\documentclass[fleqn,usenatbib]{mnras}

\usepackage{newtxtext,newtxmath}

\usepackage[T1]{fontenc}

\DeclareRobustCommand{\VAN}[3]{#2}
\let\VANthebibliography\thebibliography
\def\thebibliography{\DeclareRobustCommand{\VAN}[3]{##3}\VANthebibliography}

\usepackage{graphicx}	
\usepackage{amsmath}	
\usepackage{amssymb}

\title[Local Group mass estimators from cosmological simulations]{Local Group timing argument and virial theorem mass estimators from cosmological simulations}

\author[O. V. Hartl \& L. E. Strigari]{
Odelia V. Hartl,$^{1,2}$\thanks{E-mail: odeliah@tamu.edu}
Louis E. Strigari$^{2}$
\\
$^{1}$Department of Physics,
University of Oregon, Eugene, OR, 97403 USA\\
$^{2}$Mitchell Institute for Fundamental Physics and Astronomy, Department of Physics and Astronomy, Texas A\&M University, College Station, TX, USA 77843
\\
}

\date{Accepted 2022 February 4. Received 2022 February 4; in original form 2021 July 27}

\pubyear{2021}

\begin{document}
\label{firstpage}
\pagerange{\pageref{firstpage}--\pageref{lastpage}}
\maketitle

\begin{abstract}
We identify Local Group (LG) analogs in the IllustrisTNG cosmological simulation, and use these to study two mass estimators for the LG: one based on the timing argument (TA) and one based on the virial theorem (VT). Including updated measurements of the Milky Way-M31 tangential velocity and the cosmological constant, we show that the TA mass estimator slightly overestimates the true median LG-mass, though the ratio of the TA to the true mass is consistent at the approximate 90\% c.l. These are in broad agreement with previous results using dark matter-only simulations. We show that the VT estimator better estimates the true LG-mass, though there is a larger scatter in the virial mass to true mass ratio relative to the corresponding ratio for the TA. We attribute the broader scatter in the VT estimator to several factors, including the predominantly radial orbits for LG satellite galaxies, which differs from the VT assumption of isotropic orbits. With the systematic uncertainties we derive, the updated measurements of the LG mass at 90\% c.l. are $4.75_{-2.41}^{+2.22} \times 10^{12}$ M$_\odot$ from the TA and $2.0_{-1.5}^{+2.1} \times 10^{12}$ M$_\odot$ from the VT. We consider the LMC's effect on the TA and VT LG mass estimates, and do not find exact LMC-MW-M31 analogues in the Illustris simulations. However, in LG simulations with satellite companions as massive as the LMC we find that the effect on the TA and VT estimators is small, though we need further studies on a larger sample of LMC-MW-M31 systems to confirm these results.
\end{abstract}


\section{Introduction}
\label{sec:Intro}
The Milky Way (MW) and M31 and their associated dark matter (DM) halos are the dominant mass components of the Local Group (LG). The kinematics of stars and satellite galaxies bound to these galaxies provide an estimate of the DM halo masses of the MW~\citep{2013ApJ...768..140B,2017MNRAS.468.3428P,2019MNRAS.484.5453C,2020MNRAS.494.4291C,2020MNRAS.494.5178F} and M31~\citep{2012ApJ...753....8V}. These analyses show that M31 is likely more massive than the MW, and that the total mass in these systems is  $M_{MW} + M_{M31} \simeq [2-3] \times 10^{12}$ M$_\odot$.

In addition to the MW and M31, there are dozens of dwarf galaxies in the LG~\citep{2012AJ....144....4M,2019ARA&A..57..375S}, the most massive of which, the Large Magellanic Cloud (LMC) and M33, being $\lesssim 20\%$ of the MW/M31 mass~\citep{2016MNRAS.456L..54P,2019MNRAS.487.2685E}. Some of these dwarf galaxies are satellites that are bound to either the MW or M31, while others are bound to the LG system. In sum, the dwarf galaxies represent a small fraction of the total LG mass. 

\par The dynamics of the MW/M31 and the associated dwarf galaxies encode important information on the LG. The ``timing argument" (TA) is straightforward and often-used dynamical model to estimate the LG mass, $M_{LG}$. The TA assumes that the MW and M31 have broken off from the cosmic expansion and they can be modelled as a system on a two-body Keplerian orbit~\citep{1959ApJ...130..705K}. Applications of the TA, including both the radial and tangential velocities of M31, find $M_{LG} \sim 5 \times 10^{12}$ M$_\odot$~\citep{2008ApJ...678..187V,2012ApJ...753....8V}. TA mass estimates tend to be larger than the sum $M_{MW} + M_{M31}$ as deduced from satellite kinematics, and the origin of this discrepancy has been subject of several recent studies, most of which investigate possible systematic uncertainties in the LG mass as determined from the TA. For example, the TA mass is sensitive to the local circular speed of the Sun, and may be affected by the LMC~\citep{2016MNRAS.456L..54P}.~\cite{2013MNRAS.436L..45P} and~\citet{2014MNRAS.443.2204P} find that including a cosmological constant term in the equations of motion increases the deduced LG mass by $\sim 13\%$. More generally, the TA mass depends on the cosmological model of dark energy~\citep{2020JCAP...09..056M}, possible modifications to gravity~\citep{Benisty:2020kys}, and possible past encounters which manifest in the form of additional angular momentum in the LG~~\citep{2019arXiv190403153B, 2021A&A...656A.129B}. 

Cosmological simulations may also be used to estimate the LG mass. Identifying LG-like systems in the dark matter-only Millennium simulation,~\cite{2008MNRAS.384.1459L} showed that the TA-mass estimator is largely unbiased. These authors provide an estimate of the intrinsic (or ``cosmic") scatter that must be accounted for when deducing $M_{LG}$ from the TA.~\citet{2014ApJ...793...91G} use the dark matter-only Bolshoi cosmological simulation and include the impact of the MW/M31 tangential velocity and environmental constraints and show that the TA overestimates the true mass by a factor $\sim 1.3-1.6$. The LG mass may also be extracted from simulations using likelihood analyses. For given cuts on the velocity, separation, and environmental properties, including both the tangential and radial velocities, and depending on the specific method, these estimates span a range $M_{LG} \sim [1-5] \times 10^{12}$ M$_\odot$~\citep{2014ApJ...793...91G,2017JCAP...12..034M,2017MNRAS.465.4886C,2021PhRvD.103b3009L}.~\citet{2020ApJ...890...27Z} use simulations with semi-analytic models to deduce a LG mass at the higher end of this range. This range of LG masses is also consistent with that obtained by~\citet{2013ApJ...775..102P} using a numerical least action method.

Given the large spread in the LG mass as determined from the TA and from simulations, it is important to consider independent estimates of the LG mass. Such an additional LG mass estimate comes from the application of the Virial Theorem (VT) to the MW, M31, and the dwarf galaxies in the LG.~\cite{1999AJ....118..337C} first considered this method, starting with the assumption that the kinematics of LG galaxies may be described by the VT, finding that $M_{LG} \sim 2 \times 10^{12}$ M$_\odot$. Application of this method with a larger sample of galaxies by~\cite{2014MNRAS.443.1688D} corroborate this result. This value for $M_{LG}$ as determined from the VT is also consistent with the corresponding value deduced from examination of the properties of the Hubble Flow~\citep{2009MNRAS.393.1265K}. 

\par Motivated by understanding the systematics in how the LG mass is deduced from cosmological simulations, in this paper we identify LG-like systems in the large volume Illustris simulation~\citep{2014Natur.509..177V}. We select systems based on a series of criteria, including MW-M31 separation, relative velocity, absolute magnitudes, and density of local environment. With this sample, we provide an updated calibration of the TA mass estimator, and we determine the first estimate of the cosmic scatter in $M_{LG}$ from the VT. With Illustris, we are able to provide an estimate of both the TA and the VT-deduced masses for the LG using a hydrodynamical simulation.

\par In addition to the determination of the LG mass, we use our sample to study the kinematics of low-mass dwarf satellite galaxies in the LG that are not bound to the MW or M31. We determine the ratio of the radial to the tangential velocity dispersions for this satellite population, and find that for many LG-like systems these galaxies are on preferentially radial orbits. This result has important implications for both mass estimates of the LG and for understanding its formation history. 

This paper is organized as follows. In section~\ref{sec:selection} we give a brief description of the Illustris simulation, and the cuts implemented to the simulation to gather our sample of LG analogs. Section~\ref{sec:estimates} discusses the two methods used to estimate the mass of the mock LGs, the TA and the VT. Section~\ref{sec:results} presents the results of our analysis, and in section~\ref{sec:LMC} we analyze how our results are impacted by the presence of LMC-like galaxies. Finally, in section~\ref{sec:discussion} we discuss the implications of our results. 

\section{Selection of Local Groups}
\label{sec:selection}
In this section we outline our method for identifying LG-like systems in the Illustris simulation. Our selection criteria is based on the properties of the two most massive galaxies in the system. Within this sample, we characterize the population of fainter galaxies bound within the system, focusing specifically on the number of fainter galaxies and their kinematics.  

\subsection{Galaxies and halos in the Illustris Simulation}
IllustrisTNG \citep{2019ComAC...6....2N,2018MNRAS.475..624N, 2018MNRAS.475..648P, 2018MNRAS.475..676S, 2018MNRAS.477.1206N, 2018MNRAS.480.5113M} is a suite of cosmological gravo-magnetohydrodynamic simulations which follows the evolution of dark matter, cosmic gas, luminous stars, and supermassive black holes within the centers of galaxies. Within this suite, the TNG300-1 is the simulation with the largest number of particles, $N=2500^{3}$, with a dark matter particle mass of $m_{dm}=4.0 \times 10^{7}$ M$_\odot h^{-1}$ and a baryonic particle mass of $m_{baryon}=7.6 \times 10^{6}$ M$_\odot h^{-1}$. The TNG300-1 simulation is evolved in the large comoving volume of $205 \, h^{-1} {\rm Mpc}$. The initial conditions, in agreement with Planck 2015 results~\citep{Aghanim:2015xee}, are determined by the cosmological parameters: $\Omega_{(\Lambda,0)}=0.6911$, $\Omega_{(m,0)}=0.3089$, $\Omega_{(b,0)}=0.0486$, $\sigma_{8}=0.8159$, $n_{s}=0.9667$, and $h=0.6774.$ This yields an age of the Universe of 13.8 Gyr.

\par The Illustris database provides the properties of the galaxies and dark matter halos identified in 100 snapshots each associated with a unique redshift. Here we use the SubLink algorithm \citep{2015MNRAS.449...49R} to access the data associated with different merger trees. To identify LG-like systems, we are interested in the following kinematic galaxy and halo properties: 1) The position of the center of mass of each halo, which Illustris reports as the spatial position within the halo volume for the particle with the lowest gravitational potential; 2) the velocity, which is calculated as the sum of all particles velocity weighted by mass; 3) the maximum circular velocity, $V_{max}$, which is calculated from the spherically-averaged rotation curve; and 4) the mass $M_{halo}$ for each halo, which is the sum of all particle masses bound to the subhalo. Note, $M_{halo}$ does not include particle masses bound to subhalos of this halo, and differs from mass definitions used in previous related studies~\citep{2017MNRAS.468.1300P}. 

\par In addition to these kinematic properties, we extract the absolute magnitudes for each galaxy. IllustrisTNG calculates the absolute magnitude as the sum of the luminosity for each stellar particle of the halo. IllustrisTNG reports these magnitudes in eight bands; here we utilize the B-band magnitude and denote this as $M_{B}$. 

\subsection{The Local Group Sample}
In the $z=0$ snapshot, the subfind algorithm identifies $14.5 \times 10^{6}$ dark matter halos in TNG300-1, with the vast majority of these halos hosting galaxies. From this sample of halos, we choose systems that best resemble the LG, based on the following criteria. Our cuts and the corresponding sample sizes subsequent to each cut are summarized in Table~\ref{tab:LG_sample}. 

\begin{table}
	\centering
	\caption{To identify mock-Local Groups in Illustris, this table records the cuts imposed on the 97,574 individual galaxies with a magnitude $-22.3<M_{B}<-19.3$ starting by identifying pairs as those with a separation distance $|\Vec{r}|$. $V_{r}$ represents the total relative radial velocity between the two galaxies in the pair, where a negative value implies approaching pairs. $V_{max}$ is the maximum circular velocity of the galaxy, where we impose an upper bound to remove the most massive galaxies from the sample.}
	\label{tab:LG_sample}
	\begin{tabular}{lcr} 
	    \hline
		\hline
		Cut & & pairs \\
		\hline
    $500$ kpc $\leq |\Vec{r}| \leq 1$ Mpc & & 4133\\
    Isolation & & 798\\
    $V_{r}<0$ & & 658\\
    $V_{max}<300$\,km\,s$^{-1}$ & & 613\\
		\hline
	\end{tabular}
\end{table}

We begin by identifying galaxies with B-band magnitude similar to that of the MW and M31. For the MW the estimated B-band magnitude is $M_B = -20.9$, while for M31 it is $M_B = -20.7$~\citep{1999A&ARv...9..273V}. To include a sufficiently large number of pairs in our sample, we expand this range and select galaxies with B-band magnitude within $-22.3 < M_{B} < -19.3$. This cut results in a sample of 97,574 galaxies. 

From this sample of galaxies chosen based on absolute magnitude, we add a cut to restrict to pairs of galaxies that have a separation similar to that of MW and M31. The distance between the MW and M31 is $D=783\pm25$ kpc ~\citep{2012AJ....144....4M}. Cutting on galaxy pairs within this observed range would result in an unreasonably small sample of LG systems. In order to obtain a sufficiently large sample size of LG pairs, we take as our distance cut pairs of galaxies with a separation $500$ kpc $< |\Vec{r}| < 1000$ kpc. The lower limit of 500 kpc is motivated to ensure that the dark matter halos of the two pairs galaxies do not overlap, while the upper limit of $1000$ kpc is motivated to ensure that the galaxies are bound and are detached from the Hubble Flow. 

Even after performing the above distance cut, we must be careful to ensure that there is not a third galaxy, with magnitude comparable (i.e. within $-22.3 < M_{B} < -19.3$) to the MW or M31, in close proximity to the identified pair. To account for this possibility, we keep only those pairs that have two unique halo ids, so that each member of the pair is not also assigned to a third pair. With this additional cut, our combined sample after absolute magnitude and distance cuts contains 4133 pairs. 

To compare our magnitude cut to previous cuts that consider stellar masses of galaxies, ~\citet{2020ApJ...890...27Z} found 12 pairs in Illustris TNG100-1 using a stellar mass cut of $4.0 \times 10^{10} \, \rm{M}_{\odot} < \rm{M}_{MW} < 8.0 \times 10^{10} \, \rm{M}_{\odot}$ for the Milky Way mass, and $8.0 \times 10^{10} \,  \rm{M}_{\odot} < \rm{M}_{M31} < 13.0 \times 10^{10} \, \rm{M}_{\odot}$ for M31 mass. In this work, they also use a slightly stricter distance cut of $600$ kpc $< | \vec{r} | < 1000$ kpc. When we perform these stellar mass cuts and distance cuts to Illustris TNG300-1, we find 427 pairs. Accounting for the difference between the simulation volumes of TNG100-1 and TNG300-1, the sample of LGs that we obtain is consistent with that of ~\citet{2020ApJ...890...27Z}.

For our next cut, we ensure that each pair is sufficiently isolated from another massive system that would gravitationally perturb the LG pair. For this cut, we remove pairs if a galaxy with $M_B < -19.3$ is found within a distance of $3$ Mpc from the barycenter of the pair, or if a dark matter halo with $V_{max}>150$\,km\,s$^{-1}$ is found within  $1$ Mpc of the barycenter of the pair. This isolation criteria is similar to that of~\citet{2016MNRAS.457..844F}, who use an isolation cut of 2.5 Mpc. These cuts are implemented to identify pairs with a similar local environment to the observed LG. Our magnitude cut is motivated by demanding that nearby galaxies are fainter than the most luminous nearby galaxy outside of the LG, Centaurus A, which is $\sim$ 3 Mpc from the LG. The $V_{max}$ cut ensures the pair dominates the gravitational potential within its local volume, and thus dominates the dynamics of the LG-like system. After this isolation cut we are left with 798 isolated pairs.

The MW and M31 are observed to approach one another with a velocity (in the Galactocentric frame) of $ V_{r}=-109.3 \pm 4.4 \, {\rm km} \, {\rm s}^{-1}$. The bound on the tangential velocity from HST observations is $ V_{t} < 34$ km/s~\citep{2012ApJ...753....8V}, while recent measurements from Gaia DR2~\citep{2019ApJ...872...24V} and EDR3~\citep{2021AJ....161...58F} indicate larger mean values of $V_{t} \simeq 79$ km/s. All these measurements are still consistent with the observed MW-M31 system being on a nearly radial orbit. To match this kinematic criteria with our sample of pairs, we calculate the total relative radial velocity, $V_{r}$, for the two main galaxies of each pair. Since the simulation provides peculiar velocities, we add in the expansion velocity to obtain the total relative radial velocity for the pair. If we cut our above sample to simply ensure that the galaxies are approaching with $V_{r} < 0 \, {\rm km} \, {\rm s}^{-1}$, on top of the above magnitude, distance, and isolation cuts we are left with a sample of 658 pairs. In the analysis below we examine the impact of restricting the range of allowed radial velocities even further. 

\begin{figure}
	\includegraphics[width=\columnwidth]{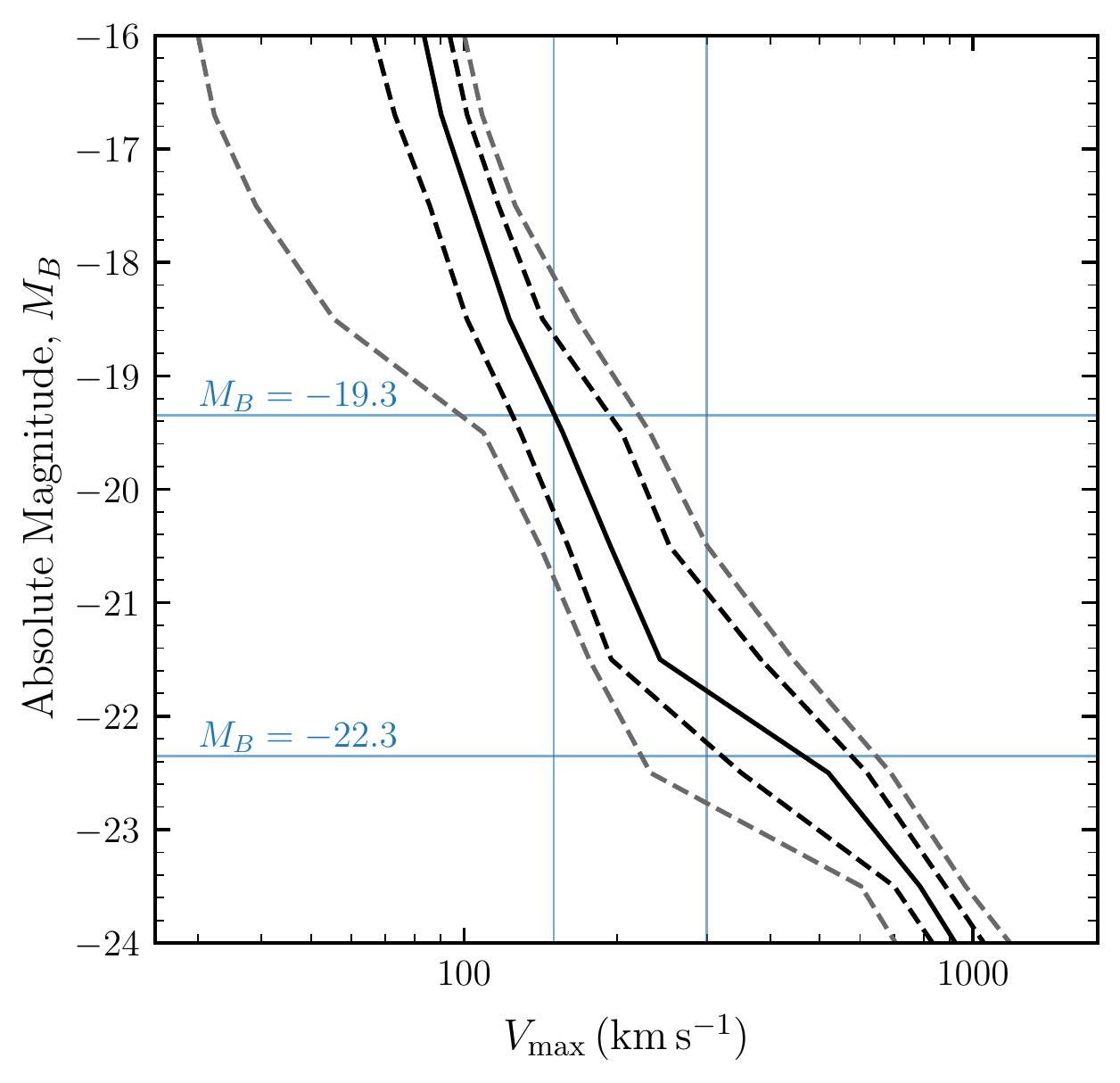}
    \caption{Absolute magnitude versus maximum circular velocity for galaxies in the Illustris TNG300-1 simulation. We identify 416,145 halos with a magnitude $M_{B}<-16$. The solid black curve indicates the median value, the black-dashed curves indicate the 16\% and 84\% points from the median, and the gray-dashed curves indicate the 5\% and 95\% points from the median. The blue vertical lines represent the $V_{max}$ range of $150$\,km\,s$^{-1}\leq V_{max} < 300$\,km\,s$^{-1}$ used in~\citet{2008MNRAS.384.1459L}. The horizontal lines indicate the magnitude range of $-22.3 < M_{B} < -19.3$, chosen as our initial cut.}
    \label{fig:vmax_mag_Illustris}
\end{figure}

Note that in choosing our sample based on absolute magnitude, we have neglected the impact of the dark matter halo mass, or similarly the maximum circular velocity, in our cuts. To better understand the range of maximum circular velocities that our absolute magnitude cuts correspond to, in Figure~\ref{fig:vmax_mag_Illustris} we show the relation between absolute magnitude and maximum circular velocity. As indicated, there is a significant scatter in $V_{max}$ over the absolute magnitude range that we consider. To remove galaxies with $V_{max}$ much larger than that of the MW or M31 galaxies, we add the constraint $V_{max}<300$\,km\,s$^{-1}$ on top of our above cuts, which results in a sample 613 galaxy pairs. 

Our cut in $V_{max}$ is similar to that implemented in~\cite{2008MNRAS.384.1459L}, who used the Millenium simulation to identify LG-like systems based on the range $150$\,km\,s$^{-1}\leq V_{max} < 300$\,km\,s$^{-1}$ as their primary sample. Examining Figure~\ref{fig:vmax_mag_Illustris} suggests that our initial magnitude cut above  corresponds to a range in $V_{max}$ that is larger than that considered in~\cite{2008MNRAS.384.1459L}, and does indeed introduce unreasonably massive halos in our sample. An additional benefit of choosing our sample based on magnitude is that it allows us to compare to the results of~\cite{2008MNRAS.384.1459L}, who use halo mass as their primary criteria.  

\subsection{Outer Local Group Members}

In addition to the two massive galaxies, we will be interested in the fainter galaxies that are gravitationallly-bound to the LG systems. In particular, we are interested in the subsample of galaxies that are bound to the LG system, but are not bound to either of the two massive galaxies. We refer to these galaxies as outer LG members (OLGM). 

From our LG sample identified above, we consider two cuts to extract OLGMs: first a cut on the absolute magnitude, and second a cut on the dark matter mass. For an absolute magnitude limit, we take the limit on the B-band magnitude to be $M_{B}<-10$, which approximately corresponds to the faintest satellite in the observed LG that is not bound to either the MW or to M31~\citep{2012AJ....144....4M,2019ARA&A..57..375S}. For our dark matter halo cut, we take the lower bounds on the dark matter halo mass of $M_{dm} > 10^{10} \, {\rm M}_\odot$ and $M_{dm} > 10^9 \, {\rm M}_\odot$. This corresponds to a cut on systems with dark matter mass approximately larger than the most massive luminous satellite galaxies of the MW and M31. 

On top of these mass and luminosity cuts, 
we must be careful to identify OLGMs that are bound to either of the two massive galaxies or are too far out into the Hubble flow. For the former, we tag the specific OLGMs that are within $< 350$ kpc from either of the two most massive galaxies. This cut tags galaxies that are within the approximate virial radius of the MW and M31. For the latter, we exclude OLGMs that are $> 1.5$ Mpc from either of the two massive galaxies, thereby excluding systems that are strongly affected by the Hubble flow. Though we only utilize the $> 1.5$ Mpc cut for our fiducial analysis below, we do compare to the results when the $< 350$ kpc cut is also implemented and find a negligible difference. Note that the $> 350$ kpc and $< 1.5$ Mpc distance cut is similar to that used in the observational sample  in~\citet{2014MNRAS.443.1688D}.


\begin{figure*}
    \includegraphics[width=0.47\textwidth]{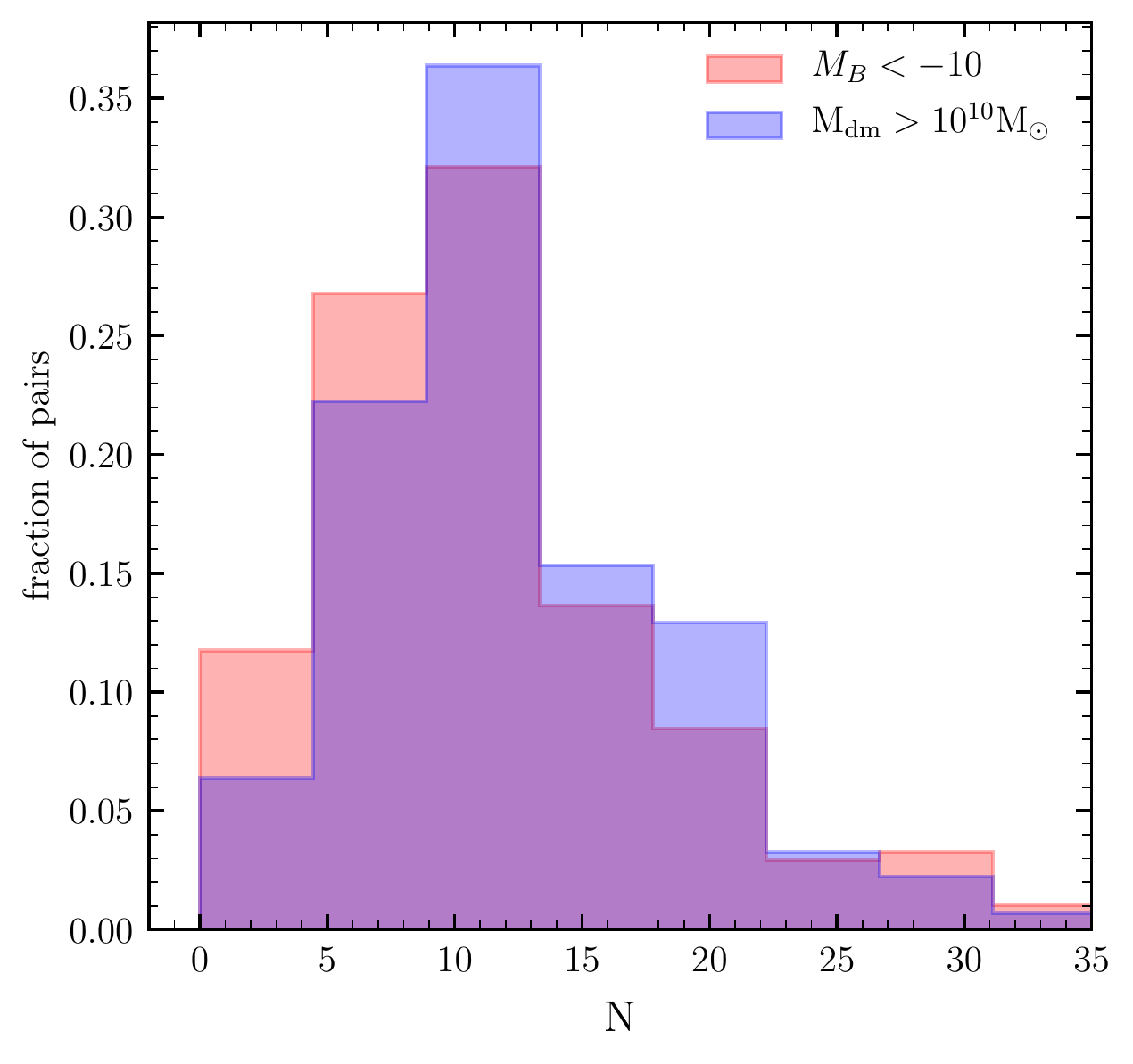}
	\includegraphics[width=0.46\textwidth]{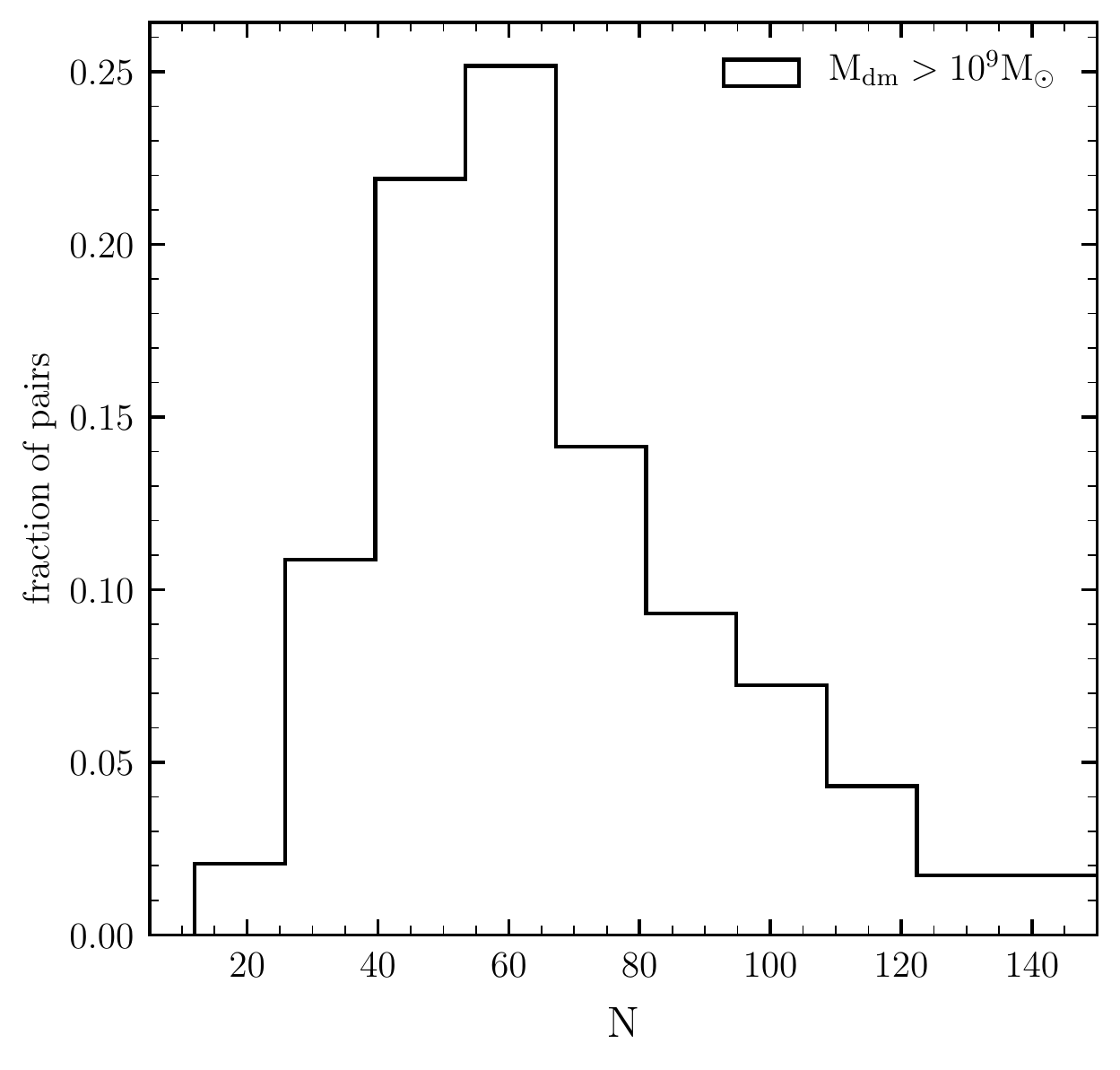}
    \caption{Distribution of $N$, the number of outer Local Group members per Local Group, found within $  D < 1.5 \, {\rm Mpc}$. The left plot shows this distribution for the luminosity cut sample in shaded red and the dark matter halo mass sample $> 10^{10} \, \rm{M}_{\odot}$ in shaded blue. The region where these distributions overlap appears in purple. The right plot shows the distribution of $N$ for the relaxed dark matter halo mass cut sample of $> 10^{9} \, \rm{M}_{\odot}$.}
    \label{fig: Num_350_Olgm}
\end{figure*}

For OLGMs identified with the distance cut of $D <1.5 $ Mpc, Figure \ref{fig: Num_350_Olgm} shows the distribution of $N$. The left plot shows the distribution of $N$ for the OLGM luminosity cut, $M_{B} <-10$, and the strict OLGM dark matter mass cut of $M_{dm} > 10^{10}$ M$_{\odot}$. The right plot shows the distribution of $N$ for the relaxed dark matter mass cut of $M_{dm} > 10^{9}$ M$_{\odot}$. The two distributions on the left plot are similar, though the sample cut on dark matter halo mass $> 10^{10}$ M$_\odot$ has a larger mean of $\langle N \rangle = 12.4$, compared to the mean of $\langle N \rangle = 11.4$ for the luminosity sample cut. Reducing the lower limit on dark matter halo mass to $> 10^9$ M$_\odot$ yields a mean of $\langle N \rangle = 67.5$ OLGMs per LG. When initially requiring each OLGM to be energetically bound, the number of OLGMs associated with each pair decreases. This reduces the LG sample size when we require $N>0$. This sample yields a similar behavior with a mean of $\langle N \rangle =8.3$ for the luminosity sample, $\langle N \rangle = 7.9$ for the $> 10^{10}$ M$_\odot$ dark matter halo mass cut, and $\langle N \rangle = 56.4$ for the relaxed $> 10^{9}$ M$_\odot$ dark matter halo mass cut. 

We impose two final cuts to obtain our sample of mock-LG systems which we will use as our sample to study the VT mass estimator below. First, we require at least 10 OLGMs per LG, i.e., $N \geq 10$. Requiring a minimum number of OLGMs is motivated by the number of OLGMs observed in the LG, $N=17$. Second, we place a restriction on the mass ratio $m =M_{M31}/M_{MW}$. We take the constraint on this quantity from ~\citet{2014MNRAS.443.1688D}, who used outer LG kinematics to estimate $m = 2.30$. Since our choice of which galaxy in the main pair is the MW or M31 is arbitrary, we restrict this value to a range of $0.3 \leq m \leq 3.3 $. 

In addition to their number, we are also interested in characterizing the kinematics of the OLGM  sample. To examine the kinematics, we first identify the position of the barycenter,  where the barycenter is calculated as the center of mass of the two massive galaxies in each system. We then determine the radial velocity, $V_{r,\imath}$, of each OLGM relative to the LG barycenter, excluding the two massive galaxies. The radial velocity dispersion of the OLGMs is then calculated as $\sigma_r^2 = \frac{1}{N-1} \sum_{\imath=1}^N (V_{r,\imath} - \bar V_r)^2$, where $\bar V_r$ is the mean radial velocity of the sample. This assumes that the radial velocity dispersion is constant throughout the LG, which is an appropriate assumption given the small sample sizes that we are considering. 

\begin{figure*}
	\includegraphics[width=\columnwidth]{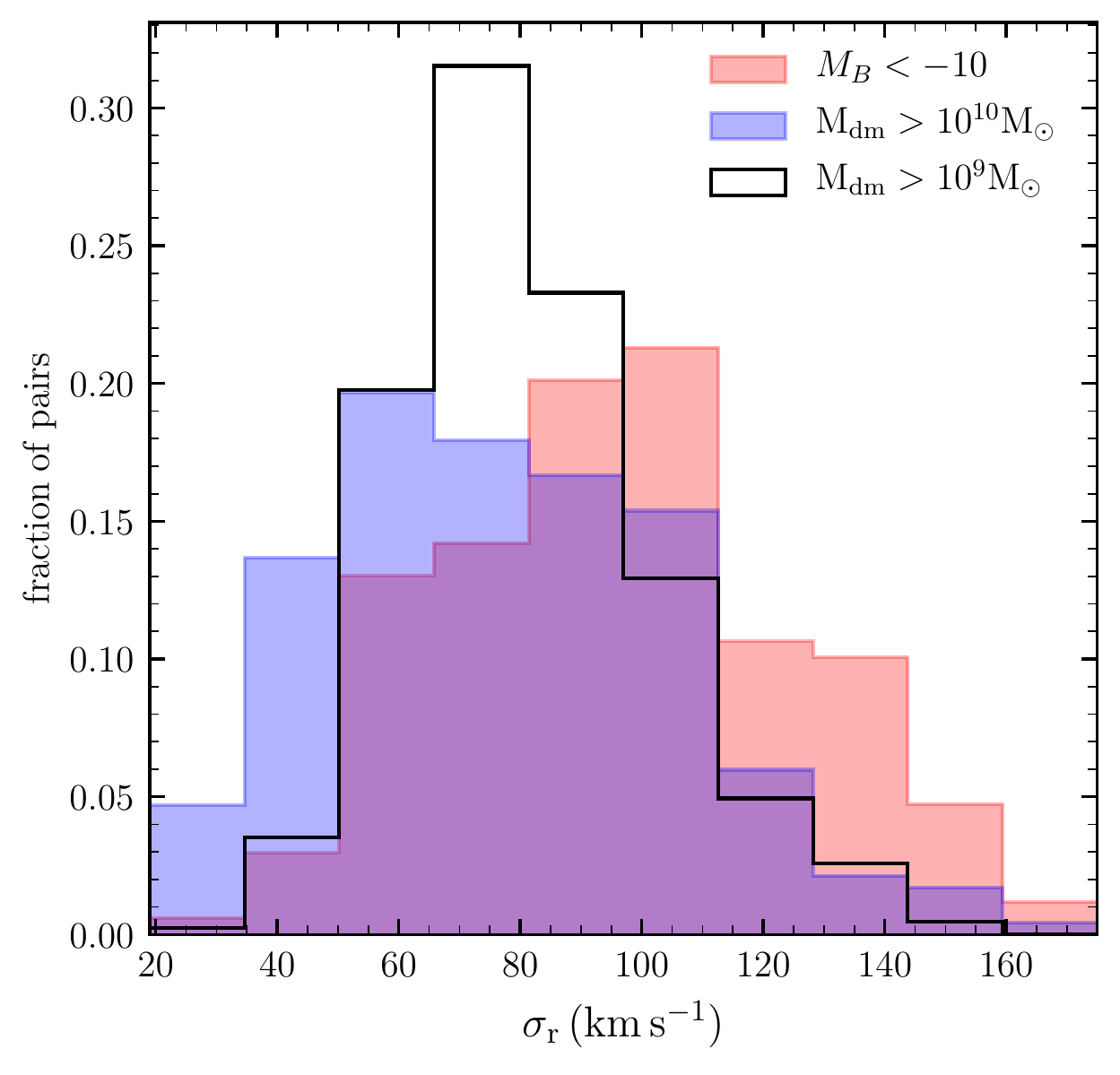} 
	\includegraphics[width=\columnwidth]{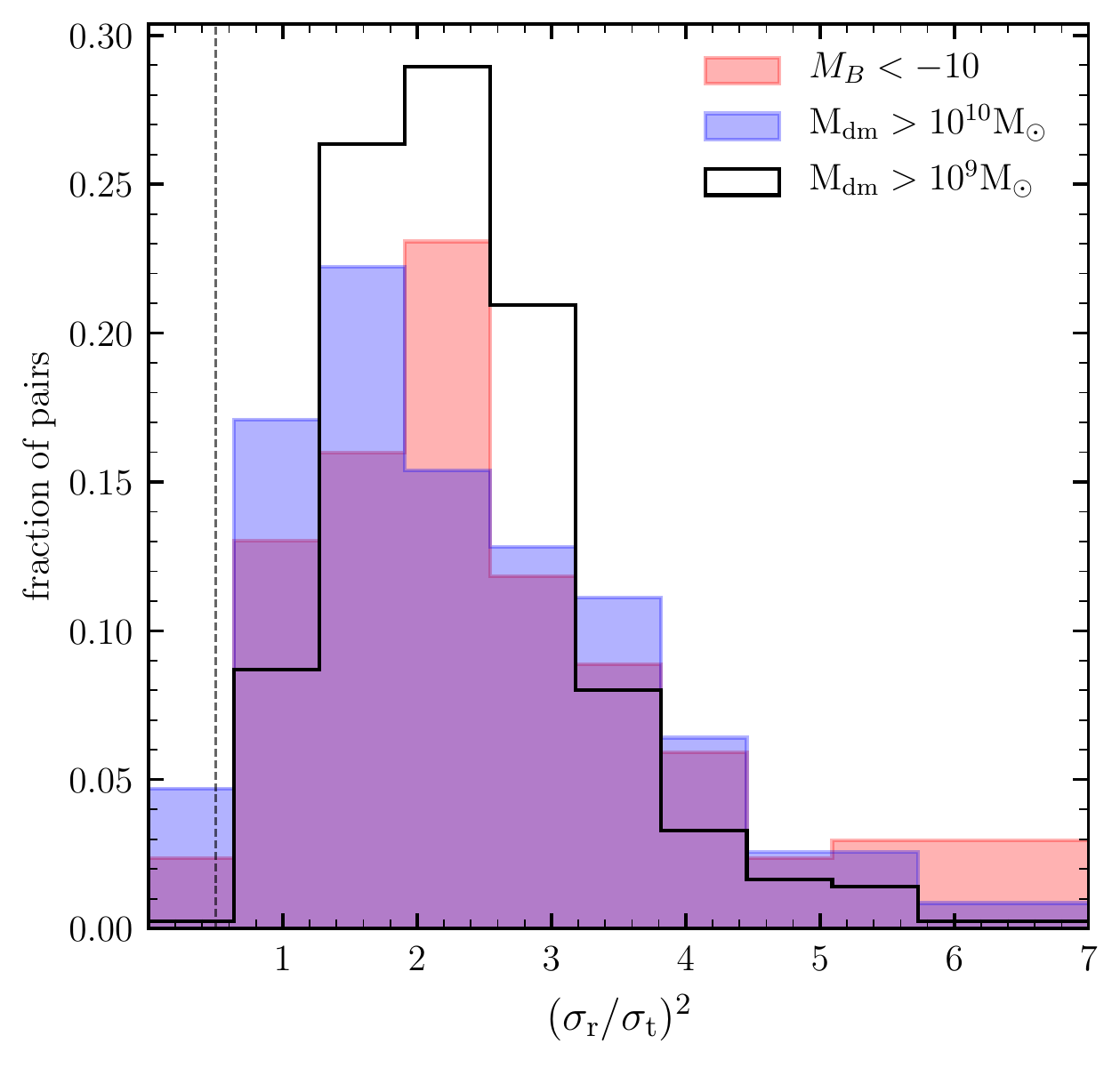} \\
    \caption{The left plot shows the distribution of the barycentric radial velocity dispersion for the $M_{B}<-10$ cut sample in shaded red, dark matter halo mass cut sample $>10^{10} \, \rm{M}_{\odot}$ in shaded blue, and dark matter halo mass cut sample $>10^{9} \, \rm{M}_{\odot}$ in black. The right plot shows the distribution of the ratio of the radial velocity dispersion to the tangential velocity dispersion squared for these same samples. For both plots, we keep only those mock-Local Group's with $N \geq 10$, and M31 to MW mass ratios 0.3$\leq m \leq $3.3. The vertical dashed-black line on the right at 0.5 indicates the isotropic value.  }
    \label{fig:sigma_350}
\end{figure*}

Figure~\ref{fig:sigma_350} (left) plots the distribution of the radial velocity dispersion for the sample with the distance cut of $D < 1.5$ Mpc and the additional cuts $N\geq 10$ and  $0.3 \leq m \leq 3.3$ described above. Implementing these cuts yields 169 pairs, 234 pairs, 425 pairs for the absolute magnitude, $M_{dm} > 10^{10} \, \rm{M}_{\odot}$, and $M_{dm} > 10^{9} \, \rm{M}_{\odot}$ cuts respectively. The histogram distribution of the radial velocity dispersion for the three samples are similar. We see the sample of mock-LG's with OLGMs identified with the luminosity cut has the largest mean radial velocity dispersion of $97.3  \, {\rm km} \, {\rm s}^{-1}$. The mean radial velocity dispersions are $79.2  \, {\rm km} \, {\rm s}^{-1}$ and $82.0  \, {\rm km} \, {\rm s}^{-1}$ for  $M_{dm} > 10^{10} \, \rm{M}_{\odot}$, and $M_{dm} > 10^{9} \, \rm{M}_{\odot}$ respectively.

The velocity dispersions we obtain for the three samples are similar, and further are consistent with the radial velocity dispersion measured from the observed LG~\citep{2014MNRAS.443.1688D}. Note that the observed velocities are heliocentric radial velocities, which must then be converted to a barycentric frame by measuring the location of the MW-M31 barycenter. While this conversion adds uncertainty to the velocity dispersion as deduced from observations, it does not add an uncertainty to our analysis given that we identify the barycenter directly from the simulation. 

  Figure~\ref{fig:sigma_350} (right) shows the corresponding ratio of the radial to the tangential velocity dispersion, $\sigma_t^2$, for each LG in the sample. The tangential velocity is defined so that for an isotropic distribution, $\sigma_r^2/\sigma_t^2 = 0.5$, where this value is denoted by the vertical line. The peak at $\sigma_r^2/\sigma_t^2 \simeq 2$ implies that OLGMs are on dominantly radial orbits, likely due to infall into the LG. Below we discuss the implications of $\sigma_r^2/\sigma_t^2$ on the equilibrium nature of the LG. 

\section{Mass estimates for Local Group} 
\label{sec:estimates}
In this section we discuss our mass estimation methods for the LG, starting with the timing argument and then moving on to the virial mass estimator. 

\subsection{Timing Argument} 
The TA has been long studied as a LG mass estimator~\citep{1959ApJ...130..705K}. The TA assumes that the MW and M31 can be described by a Keplerian orbit, with $z=0$ boundary conditions given by present-day separation, relative velocity, and the age of the Universe. From these assumptions and observations, the total LG mass $M_{LG}=M_{MW}+ M_{M31}$ may be determined. 

Implementing the TA involves solving for the evolution of the MW-M31 separation. In $\Lambda\mathrm{CDM}$, this relative separation, $\vec{r}$, is given by  
\begin{equation} 
\frac{d^{2} r}{dt^{2}}= -\frac{G M_{LG}}{r^{2}}+ H_0^2 \Omega_{(\Lambda,0)} r. 
\label{eq:MTAKEP} 
\end{equation}
Equation~\ref{eq:MTAKEP}  is solved for $r(t)$ in 3D, given the above boundary conditions and the condition that $r \rightarrow 0$ as $ t \rightarrow 0$. 

In some prior implementations of the TA, two simplifications have been invoked. The first involves neglecting the vacuum energy term in Equation~\ref{eq:MTAKEP}. A second simplification involves neglecting the tangential component of the relative motion between the MW and M31. This is a reasonable approximation, given that the magnitude of the tangential velocity is small relative to the radial velocity. In order to obtain the most accurate solution, we numerically solve  Equation~\ref{eq:MTAKEP} and account for both the vacuum energy term and the non-zero tangential velocity. 

For each mock LG in our sample, we determine the separation $|\vec{r}|$, as well as the relative velocity components $V_r$ and $V_t$. Then for a chosen $M_{LG}$, we take these as initial conditions to numerically solve Equation~\ref{eq:MTAKEP} by integrating backwards in time. The best solution for $M_{LG}$ is the one for which the separation between the two massive galaxies approaches zero as $t \rightarrow 0$ Gyr. Practically, the solution for $M_{LG}$ is obtained by extending the integration to times $t < 0$, and generating a library of $M_{LG}$ and times at which the separation goes to zero. The best value for $M_{LG}$ is obtained by interpolating between the smallest positive $t$ and the smallest negative $t$ obtained. Then given this solution for the ``timing argument mass," $M_{TA}$, we can compare to the true LG mass from the simulation to estimate the bias in the TA mass estimator. 

\subsection{Virial Theorem} 
Next we consider a LG mass estimator based on the VT. This estimator was first employed in~\citet{1999AJ....118..337C} given the sample of OLGMs known at the time, and has been developed more recently with updated data on OLGMs in~\citet{2014MNRAS.443.1688D}. 

For a system in equilibrium, the VT relates the kinetic energy $(K)$ and the potential energy $(U)$ via $2K + U = 0$. Assuming that the system is isotropic, so that the random motions are the same in each of the coordinate directions, the kinetic energy is $K = 3/2 N M_0 \sigma_{r}^2$, where $N$ is the number of galaxies in the sample, $M_0$ is the mass of each galaxy, and $\sigma_r$ is the one-dimensional radial velocity dispersion of the system. The factor of 3 arises from the assumption of isotropy. 

We take two contributions to the potential energy $U$: the gravitational contribution from the two massive galaxies, and the contribution from the cosmological constant. On the $\imath^{th}$ galaxy, the potential is then
\begin{equation}
    \phi = -G M_{MW} p_\imath -G M_{M31} q_\imath - \frac{4\pi}{3} G\rho_\Lambda |\vec r_\imath|^{2}, 
    \label{eq:phi}
\end{equation}
where $\rho_\Lambda$ is the present value of the vacuum energy density, and 
\begin{eqnarray}
p_\imath &=& \left( |\vec r_\imath - \vec r_{MW}| + a_{MW} \right )^{-1} \\
q_\imath &=& \left( |\vec r_\imath - \vec r_{M31}| + a_{M31} \right )^{-1}. 
\end{eqnarray}
Here $a_{MW}$ is the scalelength of the MW's dark matter halo, and $a_{M31}$ is the corresponding scalelength of the M31's dark matter halo. To model the DM density profiles of the MW and M31, we have followed~\citet{2014MNRAS.443.1688D} and adopted a Hernquist profile, and for the scale radii we take $a_{MW} = a_{M31} = 40$ kpc. We find that reasonable changes in the scale radius for each halo do not affect our results. 

Taking the approximation that all the satellite galaxies have equal mass, the LG mass can be estimated as
\begin{equation} 
M_{LG} = \frac{-3 N ( 1 + m ) \sigma_{r}^{2} - (8\pi/3) G \rho_\Lambda (1+m)\sum_\imath^N |\vec r_\imath|^{2}}
{G \sum_\imath^N \vec r_\imath \cdot \vec \nabla_\imath (p_\imath + m q_\imath)}, 
\label{eq:MLGVTCC} 
\end{equation} 
where $\rho_\Lambda = 0.69 \times 1.5 \times 10^7$ M$_\odot$ ${\rm Mpc}^{-3}$, and as defined above $m = M_{M31}/M_{MW}$ is the mass ratio between the MW and M31. Note that this formula is similar to that derived in~\citet{2014MNRAS.443.1688D}, the only difference being the vacuum energy term which is added to the kinetic energy term. 

With our LG sample, we implement Equation~\ref{eq:MLGVTCC} by determining $\sigma_r$ and the mass ratio $m$ for each system, and from these determine $M_{LG}$. We refer to this solution as the virial theorem mass, $M_{VT}$. Note that since $m$ is not a priori known for the MW-M31 system, we must take realistic bounds on this quantity when comparing to observations. We discuss in detail below how the assumption for $m$ affects our results.

\section{Results} 
\label{sec:results}
In this section we present the results of our comparison of the TA and the VT mass estimators to the LGs identified in the Illustris simulation. We estimate the systematics that arise from the ``cosmic scatter" in each of these measurements, and use these to provide robust estimates of the LG mass using each method. 

\subsection{Calibration of the Timing Argument mass for the Local Group}
We begin with an analysis of the TA mass estimator. Our analysis follows a similar approach to that of~\citet{2008MNRAS.384.1459L}, who utilize the dark matter-only Millenium  simulation. These authors determine the bias in the TA estimate, $M_{TA}$, as compared to the true LG mass as obtained from the simulation, where the true mass of the LG from the simulation, $M_{tr,halo}$, is defined as the sum of the MW/M31 halo analogs, $M_{tr,halo}=M_{halo,1}+M_{halo,2}$. These authors define $A \equiv M_{tr,halo}/M_{TA}$, and study the distribution of this quantity for their mock LG sample from Millenium. We use our sample of LGs to obtain a distribution of $A$, which provides an estimate of the bias in the TA mass estimator for the Illustris simulation.  

\par To get a sense of the relative importance of the different velocity cuts, we consider three subsamples of our 613 LG pairs identified above. In the first subsample, we restrict the tangential velocity to be less than the absolute value of the radial velocity, i.e., $V_{t} < |V_{r}|$. This is our loosest cut, motivated to simply ensure that the dominant component of the velocity is in the radial direction. In the second subsample, we restrict the radial velocity range to  $-195$\,km\,s$^{-1}\leq V_{r}\leq-65$\,km\,s$^{-1}$, reducing our sample to 251 pairs. This cut is motivated to span the uncertainty in the observed radial velocity between the MW and M31, and was implemented with similar motivation by~\citet{2008MNRAS.384.1459L}. For our third subsample, we consider the union of the first and second cuts. This is our smallest subsample, resulting in a total of 177 pairs.

To compare our magnitude cut to previous cuts that consider $V_{max}$, ~\cite{2008MNRAS.384.1459L}  found 11838 pairs in Millennium simulation using $150$\,km\,s$^{-1} <V_{max}<300$\,km\,s$^{-1}$ as their initial cut. For this sample of 11838 pairs, they use the same distance cut of $500$ kpc $< | \vec{r} | < 1000$ kpc, and radial velocity cut of $-195$\,km\,s $^{-1}\leq V_{r}\leq-65$\,km\,s$^{-1}$. When we perform the $V_{max}$, distance and radial velocity cuts on Illustris TNG300-1, we find 1051 pairs. Accounting for the difference in the simulation volumes, $500 \, h^{-1} {\rm Mpc}$ and $205 \, h^{-1} {\rm Mpc}$ for the Millennium and IllustrisTNG simulations respectively, our final number of mock-LG pairs is consistent with the number of pairs found by ~\cite{2008MNRAS.384.1459L}. 

\begin{table*}
	\centering
	\caption{Results from the timing argument analysis. Confidence intervals are shown for the distribution of $A$ under three differing velocity restrictions denoted in the velocity cut column.}
	\label{tab:MTA_Table}
	\begin{tabular}{llcccccr} 
		\hline
		& Velocity Cut & 5\% & 25\% & 50\% & 75\% & 95\% & \# of pairs\\
		\hline
		\hline
		& $V_{t} < |V_{r}|$ & 0.42&0.67&0.83&0.98&1.22&284\\
	& $-195$\,km\,s$^{-1}<V_{r}<-65$\,km\,s$^{-1}$& 0.39 &0.61&0.74&0.88&1.16&251\\
    &  $-195$\,km\,s$^{-1}<V_{r}<-65$\,km\,s$^{-1}$,$V_{t} < |V_{r}|$ & 0.39 &0.64&0.79&0.90&1.16&177\\
		\hline
		
	\end{tabular}
\end{table*}

\begin{figure*}
	\includegraphics[width=\columnwidth]{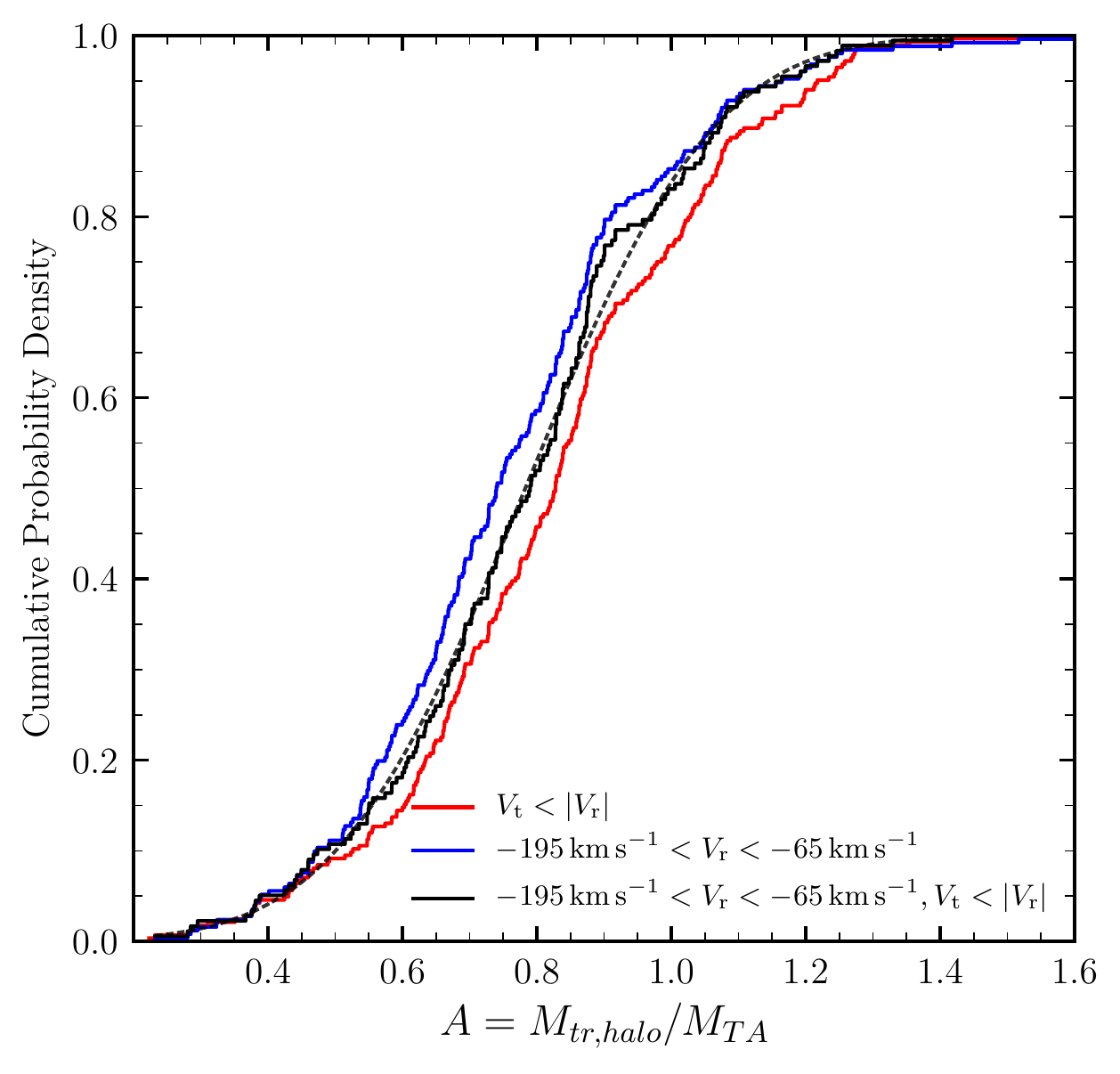}
	\includegraphics[width=\columnwidth]{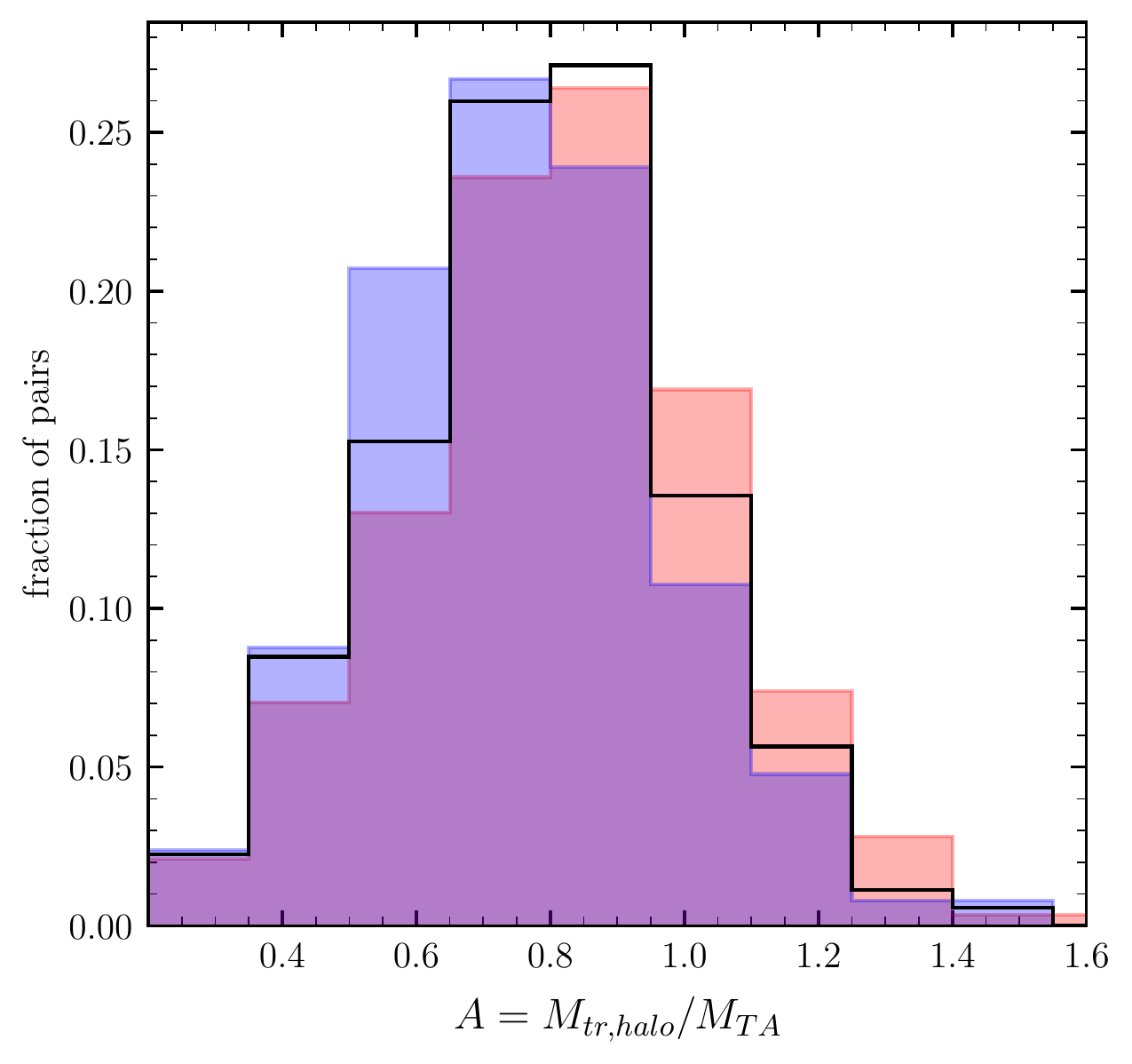}
    \caption{Cumulative probability distributions (left) and histogram distributions (right) for $A$, the ratio of the true mock-LG mass to the Timing Argument mass. We plot for three different velocity restrictions on the main sample.  First, restricting the tangential velocity to be less than the radial velocity between the main pair, $V_{t} < |V_{r}|$ (red). Second, the radial velocity restriction of $-195$\,km\,s$^{-1}<V_{r}<-65$\,km\,s$^{-1}$ (blue). Then the union of these two velocity cuts,   $-195$\,km\,s$^{-1}<V_{r}<-65$\,km\,s$^{-1}$,$V_{t} < |V_{r}|$ (black). The dashed-black curve is the best Gaussian fit for the preferred velocity sample, $-195$\,km\,s$^{-1}\leq V_{r}\leq-65$\,km\,s$^{-1}$, $V_{t} < |V_{r}|$.  }
    \label{fig:MTA_prob_dist_ls_lambda}
\end{figure*}

Figure~\ref{fig:MTA_prob_dist_ls_lambda} plots the cumulative (left) and histogram (right) distributions of $A$ for all three velocity samples. In Figure~\ref{fig:MTA_prob_dist_ls_lambda} (left) also plots the best fit gaussian curve (dashed-black) for the preferred velocity sample with $-195$\,km\,s$^{-1}\leq V_{r}\leq-65$\,km\,s$^{-1}, V_{t}< |V_{r}|$. From this we see that the distribution is very close to gaussian. In particular, comparing the mean of 0.75 to 0.77 and standard deviation of 0.22 to 0.27 for the curve fit and  Gaussian fit respectively, we see our sample has a relatively normal distribution. 

 In Table~\ref{tab:MTA_Table} we summarize our statistical results, reporting the interquartile and the 90\% range of the distributions for all three velocity cuts. The median value of $A$ and the interquartile ranges are less than unity, with $A \simeq 0.74-0.83$, depending on the specific velocity cut implemented. This implies that there is a slight bias in the TA mass estimator to predict a mass higher than the true mass obtained from the simulation. However, when considering the 90\% range about the mean, this bias is removed, with the upper bounds on the distribution in the range $A \simeq 1.16-1.22$. 
 
To understand this bias, we also calculate the timing argument mass assuming no cosmological constant, i.e. $\Lambda = 0$. This calculation yields a median value of $A$ in the range $A \simeq 0.82-0.93$. 
Thus, including the cosmological term in calculating the TA mass estimator results in a shift to a median value of $A \simeq 0.74-0.83$, indicating a larger timing argument mass estimate. This particular shift in the TA mass estimator can be understood by considering the current distance and approach velocity between the pair. When we include the cosmological term in the calculation, we account for a repulsive term in the potential. Thus, a more massive system is estimated in order to account for the current distance and approach velocity. Recall these distributions record $M_{tr,halo}/M_{TA}$, thus a higher estimate in the timing argument mass lowers the distribution.

It is interesting to compare our median values of the distribution of $A$ to that of~\citet{2008MNRAS.384.1459L}. In addition to selecting LG systems based on maximum circular velocity, these authors do not include the effect of the cosmological constant term, nor do they include the tangential velocity. Assuming $V_{t} = 0$ yields a lower bound to the TA mass estimate. In addition, excluding the cosmological term has the affect described above. Thus the combination of $V_{t} = 0$ and $\Lambda = 0$ results in their slight calculated bias in the median value of $A \simeq 0.99$. This is consistent with~\citet{2014ApJ...793...91G}, who find that including the tangential velocity overestimates the true mass by a factor $\sim 1.3-1.6$.

The different velocity cuts that we consider impact the tail of the $A$ distributions. In particular, constraining the range of the radial velocity to $-195$\,km\,s$^{-1}\leq V_{r}\leq-65$\,km\,s$^{-1}$ reduces the width of the distribution. For different velocity cuts, the distribution widens because the sample includes systems with either very large or very small radial velocities. In addition requiring the tangential velocity to be less than the radial velocity, $V_{t}<|V_{r}|$, the width of the distribution is unchanged, though the median values shift up slightly. This upward shift in the median values reflects the smaller TA mass estimate for systems with small tangential velocity.

\subsection{Timing Argument Application to Local Group}
\par We are now in position to calculate the TA mass of the LG, and to apply a correction to this estimate based on our analysis above. We assume $\Omega_{(\Lambda, 0)} =0.69$ and we utilize the observed Local Group kinematics of $V_{t} = 79 $ km s$^{-1}$, $V_{r} = -109.3$ km s$^{-1}$ and $r = 785$ kpc. 
For these assumptions, we calculate a TA mass of 
\begin{equation}
     M_{TA,LG}= 6.01 \times 10^{12} \text{ M}_{\odot}.
\end{equation}
The observational uncertainty on the TA mass is primarily due to the measured uncertainties of the radial and tangential velocities.~\citet{2012ApJ...753....8V} quote a statistical uncertainty of $0.45 \times 10^{12}$ M$_\odot$. Noting 5\% and 95\% containment points of the distribution for our preferred sample are separated by a factor of $2.97$, we see that the systematic uncertainty of the TA mass is much larger than the measured uncertainty. 

From Table~\ref{tab:MTA_Table}, the median value for the distribution of $A$ for the preferred sample of mock-LGs is 0.79. Thus, the corrected value to the timing argument mass is \begin{equation}
    M^{true}_{TA,LG}= 4.75 \times 10^{12}\text{ M}_{\odot}.
\end{equation}
The 5\% and 95\% c.l. on the $A$ distribution are $0.39$ and $1.16$ respectively. This yields a range of $2.34 \times 10^{12} \, \text{M}_{\odot} \leq M^{true}_{TA,LG}\leq 6.97 \times 10^{12}\text{ M}_{\odot}$ with $90\%$ confidence.

For comparison, assuming that $\Lambda =0$, we obtain
\begin{equation}
    M_{TA,LG}= 5.43 \times 10^{12} \text{ M}_{\odot}.
\end{equation}
For $\Lambda = 0$, the median of the $A$ distribution is $0.87$, and the 5\% and 95\% range yield values of 0.45 and 1.29, respectively. Thus, using the median value to correct the TA mass yields 
\begin{equation}
    M^{true}_{TA,LG}= 4.72 \times 10^{12}\text{ M}_{\odot}.
\end{equation}
in a range of $2.44\times 10^{12}\text{ M}_{\odot} \leq M^{true}_{TA,LG}\leq 7.01 \times 10^{12}\text{ M}_{\odot}$ with $90\%$ confidence.
Note that the 90\% confidence ranges for the $M_{true,LG}$ with and without the cosmological constant term are very similar.

\par Our finding of the mild impact of the cosmological constant in the TA mass is consistent with previous work~\citep{2013MNRAS.436L..45P}. This is because, when including the cosmological constant, the magnitude of the effective gravitational potential felt by the system is reduced relative to the case of a pure matter-only model. So for fixed observed parameters ($V_r$, $V_{tan}$, and the MW-M31 separation), the mass would have to increase to compensate for the reduced potential, and so the ratio $A$ is shifted to slightly smaller values. 

\subsection{Calibration of the Virial Method mass for the Local Group}
We now move on to an analysis of the VT mass estimator, and compare this to the LG mass derived from the simulations. Here we define the VT mass estimate as $M_{VT}$, and the ratio of the VT mass estimate to the true mass as $B \equiv M_{tr,halo}/M_{VT}$. We consider the same 613 LG-analogs, and again define $M_{tr,halo}$ as the sum of the masses of the two most massive galaxies.
For each sample, we use Eq.~\ref{eq:MLGVTCC} to calculate the mass of each mock-LG, $M_{VT}$. 
To begin, we calculate the mass ratio $m$ for each mock-LG by setting the MW and M31 masses to their true values as extracted from the simulation.

\begin{figure*}
	\includegraphics[width=0.465\textwidth]{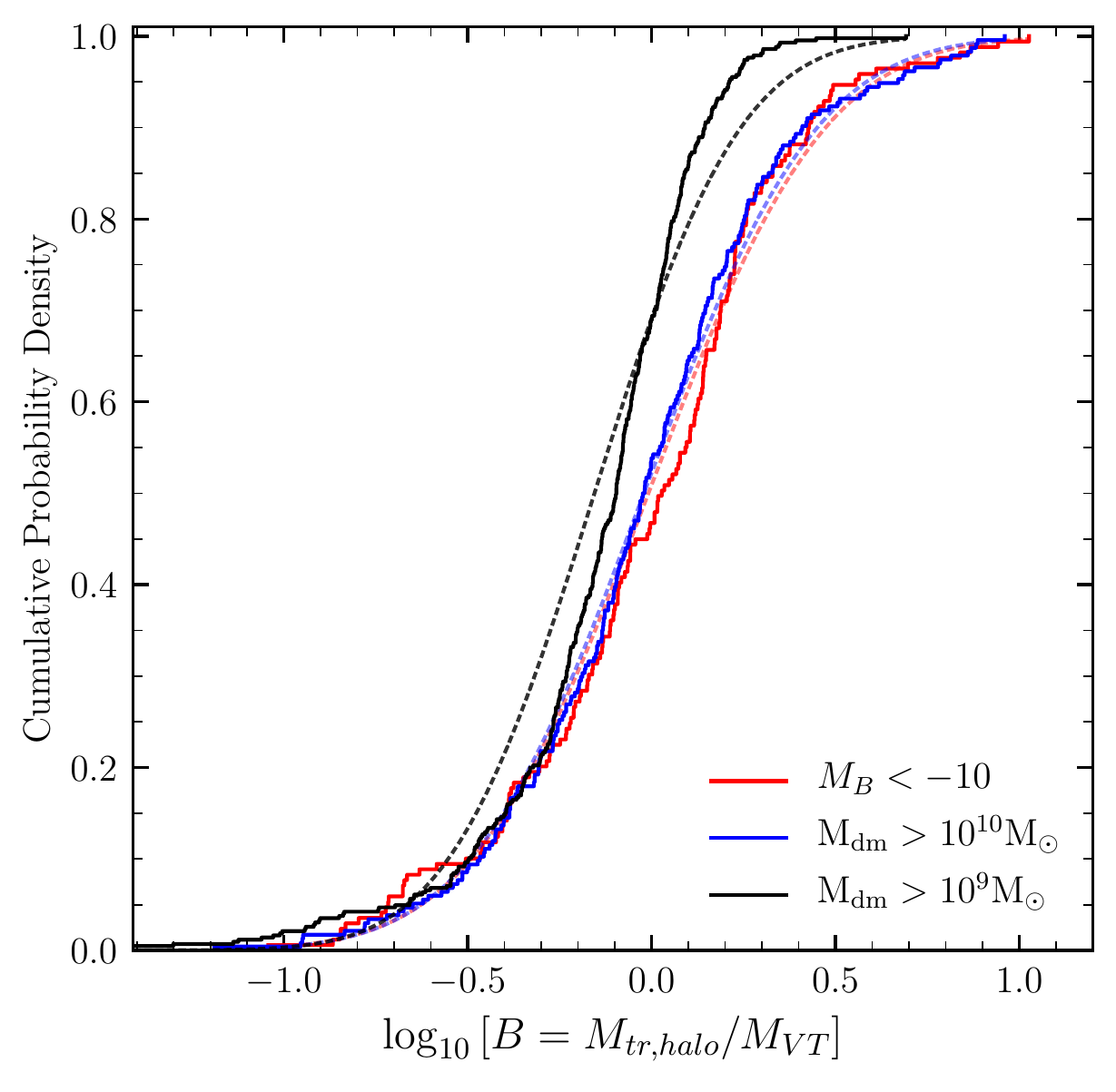} 
	\includegraphics[width=0.47\textwidth]{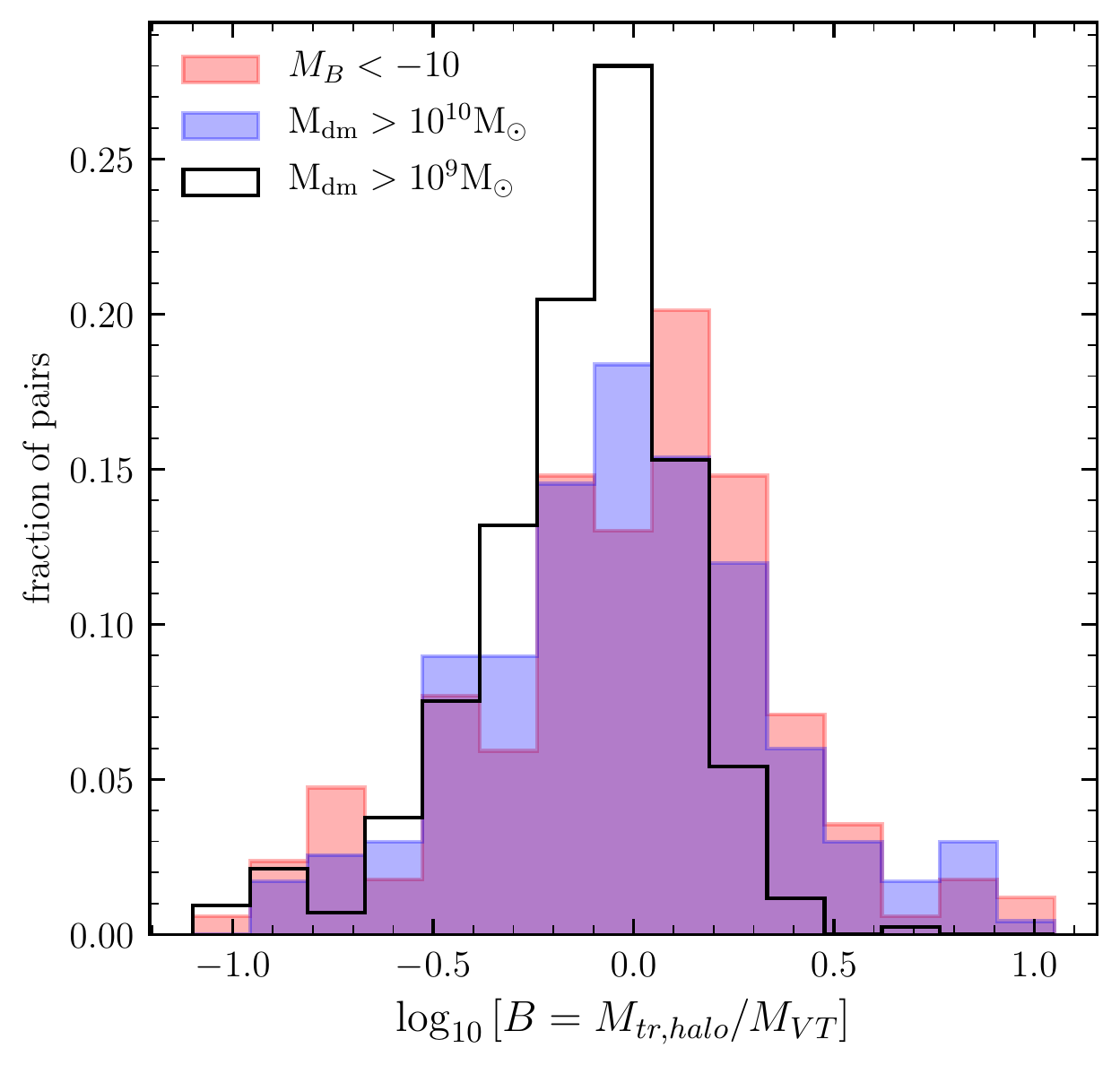} \\
    \caption{Left plot shows the cumulative distribution of $
    \log_{10} B$, the ratio of the true mass from the mock-LG to the virial theorem derived mass, for all three samples. The best fit Gaussians for each cut are shown by the dashed curves. The right plot shows the histogram distribution of $\log_{10} B$ for all three samples. }
    \label{fig:Mvt_cumdist_350}
\end{figure*}

Figure~\ref{fig:Mvt_cumdist_350} shows the distributions of $\log_{10} B$. We plot the sample with the OLGM distance criteria of $D < 1.5$ Mpc and the additional cuts, $N \geq 10$ and $ 0.3 \leq m \leq 3.3$. The left plot of Figure~\ref{fig:Mvt_cumdist_350} shows this distribution for all three samples in cumulative form. The dashed curves indicate the best fit gaussian for the three samples. The right plot of  Figure~\ref{fig:Mvt_cumdist_350} shows histogram distribution of $B$ for all three samples. 

In Table~\ref{tab:MVT_Table} we summarize our statistical results, reporting the interquartile and the 90\% points of the distributions for all three OLGM identification cuts. The median value of $B$ and the interquartile ranges are less than unity, with $B \simeq 0.80-1.07$, depending on the initial OLGM cut implemented. This implies that in general there is a slight bias in the VT mass estimator depending on the specific choice of OLGM sample. In particular in the case of our main sample, $M_{dm}>10^{9}$ M$_{\odot}$, we find a slight bias, $B = 0.8$, with a 95\% containment interval of $[0.20,1.63]$.

As noted above, current estimates of $m = M_{M31}/M_{MW}$ indicates that the mass of M31 is about a factor of two larger than that of the MW. 
This motivated an additional cut to our Local Group sample, keeping only those with $0.3 \leq m \leq 3.3$. If we instead relax this restriction on $m$, but still use the true values of their masses in the VT, there is minimal change in the $B$ distribution, yielding median values of $B \simeq 0.79-1.01$ 

When we instead consider the sample with the initial OLGM distance criteria of $350$ kpc $< D < 1.5$ Mpc and the additional cuts, $N \geq 10$ and $ 0.3 \leq m \leq 3.3$, we find a larger range for the median value between cuts of $B \simeq 0.82-1.18$. This larger range may be in part because the stricter initial OLGM distance cut yields a smaller number of pairs in our sample. 

For all of the above analysis, we chose $m$ to be the true value by extracting the true M31 and MW masses from the simulation. To check the results are not sensitive to the choice of $m$, we simply set $m=1$ in the VT, implying that the masses of the MW and M31 are equal. Note setting $m=1$ implies the barycenter is located at the midpoint of the pair, and this assumption propagates in calculating the velocity dispersion. Referring to our main sample in which we use the distance cut of $D < 1.5$ Mpc and restrict the true value of $m$ to be within $0.3 \leq m \leq 3.3$, we find the median value to be $B \simeq 0.80-1.09$. Comparing this to $B \simeq 0.80-1.07$ when we use the true value of $m$, we find the VT mass estimate is not sensitive to the value of $m$ assumed in the VT. 

\begin{table*}
	\centering
	\caption{Results from the virial theorem analysis. The first column gives the galaxy sample, the second column gives the mean number of OLGMs per LG, the third column gives the mean radial velocity dispersion per LG. Columns four through eight give the corresponding confidence intervals for each sample, with 50\% being the median of the $B$ distribution. The last column gives the number of pairs for each sample subject to the restriction $N \geq 10$. The first three rows place no restriction on the MW-M31 mass ratio, while the bottom three rows restrict the ratio to the range $0.3 \leq m \leq 3.3$.} 
	\label{tab:MVT_Table}
	\begin{tabular}{lccccccccr} 
		\hline
		OLGM cut, $D < 1.5$Mpc &$\langle N \rangle$&$ \langle \mathrm{\sigma_r\, (km \, s^{-1})}\rangle$ &5\% & 25\% & 50\% & 75\% & 95\% & \# of pairs\\
				\hline

		$M_{dm}>10^{10}\, \rm{M}_\odot$ &15.9&89.0 &0.24&0.58&0.93&1.43&3.34 &345\\
		$M_{dm}>10^{9}\, \rm{M}_\odot$ &67.9&89.8 &0.23&0.56&0.79&1.05&1.57 &569\\
        $M_{B} < -10$ & 16.2&108.8 &0.21&0.62&1.01&1.60&3.08 &274\\
		\hline
		\hline
		$0.3 \leq m \leq 3.3$\\
		\hline
		\hline
	
		$M_{dm}>10^{10}\, \rm{M}_\odot$ &14.8&79.2 &0.22&0.56&0.95&1.60&4.15 &234\\
		$M_{dm}>10^{9}\, \rm{M}_\odot$ &61.4&82.0 &0.20&0.54&0.80&1.09&1.63 &425\\
        $M_{B} < -10$ & 14.6&97.3 &0.19&0.60&1.07&1.68&3.59 &169\\
		\hline
	
	\end{tabular}
\end{table*}

\subsection{Virial Theorem Application to Local Group}
~\citet{2014MNRAS.443.1688D} have estimated the LG mass from OLGMs using the VT. These authors derive this result using radial velocity data on OLGMs within a distance of $350$ kpc $< D < 1.5$ Mpc. The result of their analysis is an LG mass estimate of 
\begin{equation}
    M_{VT,LG}= 2.5 \pm 0.4 \times 10^{12}\text{ M}_{\odot}.
    \label{eq:VTDiaz}
\end{equation}
Referring to our preferred sample with a median of $B = 0.8$ and a 90\% containment interval of 0.20-1.63, the corrected value for the LG mass estimate from the VT is then
\begin{equation}
    M^{true}_{VT,LG}= 2.0 \times 10^{12}\text{ M}_{\odot}.
    \label{eq:VTtrue}
\end{equation}
This correction was performed with our preferred sample of all OLGMs $< 1.5$ Mpc, though as noted above the $B$ distribution changes only slightly if OLGMs $< 350$ kpc are excluded. The uncertainties in Equation~\ref{eq:VTtrue} are just larger than the measured uncertainty of $0.4\times 10^{12} \text{ M}_{\odot}$ quoted in~\citet{2014MNRAS.443.1688D}. The upper and lower limits of this estimate imply a range of is $0.5\times 10^{12}\text{ M}_{\odot} \leq M^{true}_{VT,LG} \leq 4.1 \times 10^{12}\text{ M}_{\odot}$ with 90\% confidence. 
This indicates, when estimating the mass using the VT from observations, the derived value is an overestimate of the true mass, though the correction is small.

\par The fact that the VT estimator is largely unbiased, i.e. $\langle B \rangle \simeq 1$, may come as a surprise, given the simplicity of this estimator and the assumptions made in deriving it. The two key assumptions in deriving the VT are: 1) that the LG is in dynamical equilibrium, and 2) that the orbits of the OLGMs are statistically isotropic. The former assumption is most likely violated simply because the MW and M31 are falling towards one another for the first time. The application of the VT is further complicated given the fact that a large fraction of LG satellites are bound to either the MW or M31. Regarding the second assumption, this is also manifestly violated in our LG sample since the OLGMs are on largely radial orbits (Fig.~\ref{fig:sigma_350} right). If the system is assumed to be isotropic, but the underlying orbits are anisotropic, the result would be that the VT overestimates the true mass. This is indeed the trend that we observe in our distribution of $B$, albeit with a substantial scatter. 

\begin{figure*}
    \includegraphics[width=0.47\textwidth]{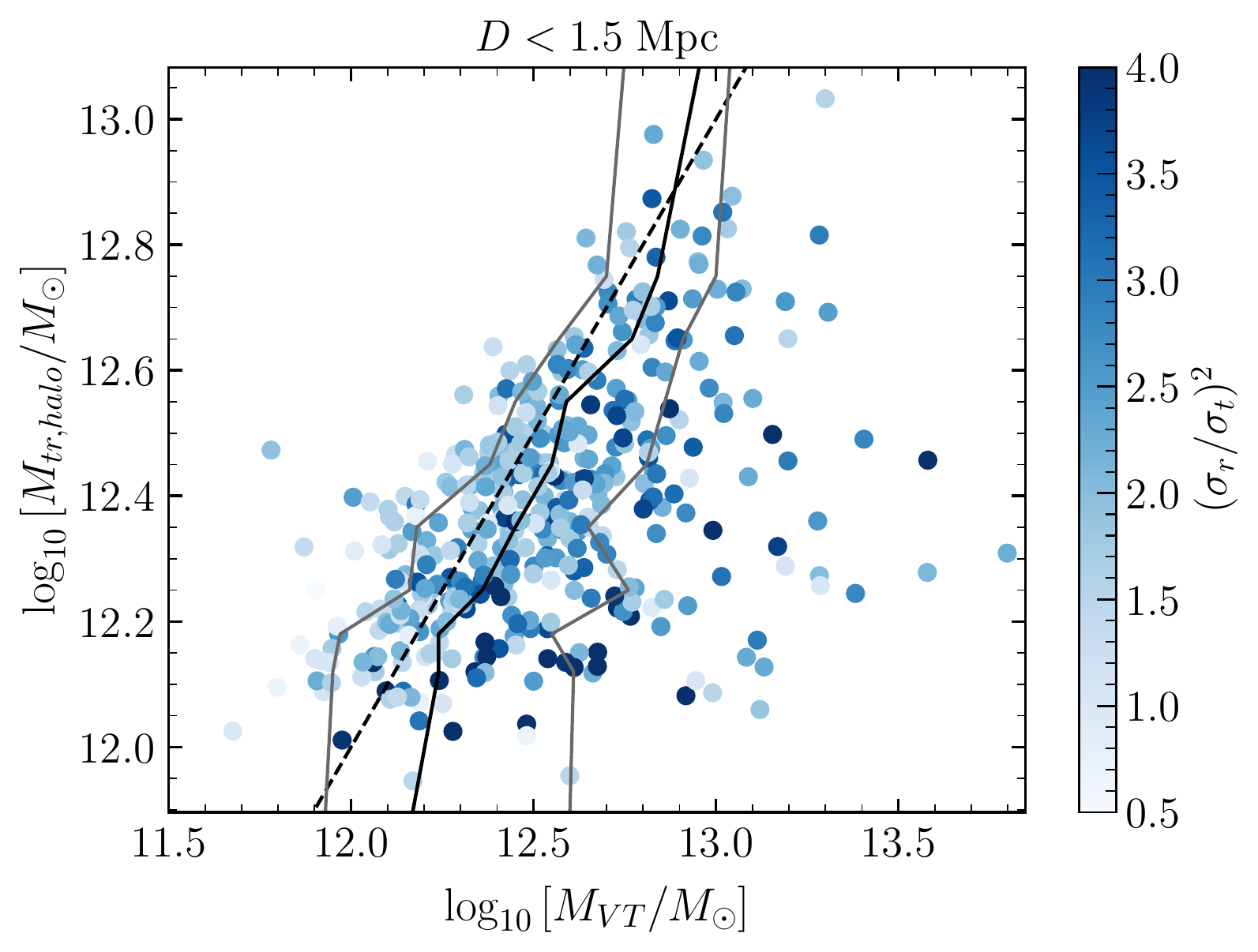}
	\includegraphics[width=0.47\textwidth]{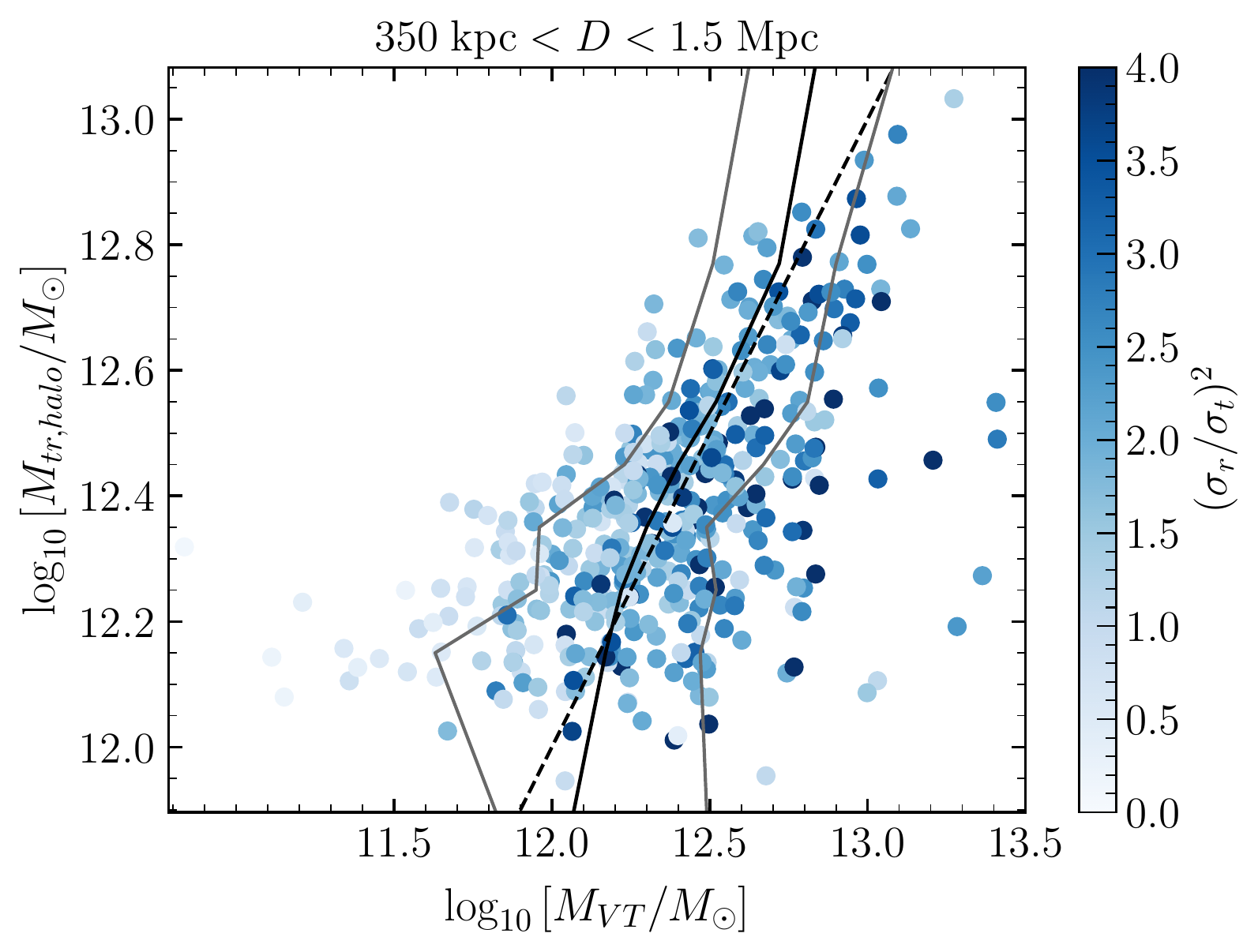}
    \caption{Log of the true mass of the mock-LG vs. the log of the mass derived from the virial theorem, $M_{VT}$. In both plots, we show the OLGM dark matter mass sample of $> 10^{9} \, \rm{M}_{\odot}$, where the left plot considers OLGM's with a distance criteria of $D < 1.5$ Mpc and the right shows the $350$ kpc $< D < 1.5$ Mpc criteria. In both plots, the dashed blacked line indicates where $M_{VT} = M_{tr,halo}$, the solid black line indicates the median of our sample, and the gray solid lines indicate the 16\% and 84\% points. The color bar indicates the value of $(\sigma_{r}/\sigma_{t})^{2}$, where the darker points represent larger values of this ratio. }
    \label{fig:VTscatter_mass}
\end{figure*}

To further examine this point, in Figure \ref{fig:VTscatter_mass}
we plot the log of the true mock-LG mass vs the log of the mass derived from the VT. We plot the OLGM dark matter mass sample of $>10^{9} \, \rm{M}_{\odot}$, where the left plot considers OLGM's with a distance of $D< 1.5$ Mpc, and the right plot considers the distance of $350$ kpc $<D< 1.5$ Mpc. In each plot the dashed-black line indicates the one-to-one line, where the true mass equals the VT mass. By taking bins of roughly $0.1$ in log of the true mass from 12.2-12.8, we calculate the median, 16\% and 84\% points of the VT mass in each of these bins. We show the median with the solid black line, and the 16\% and 84\% points with the gray lines. The color bar indicates the value of $(\sigma_{r}/\sigma_{t})^{2}$ where we have set an upper limit of 4 to cover the range on the majority of the sample. There are three points highlighted in these plots. First, in both plots, we generally see the trend of lighter points on the left and darker points on the right of the dashed-black line. This agrees with our conclusion that the anisotropic mock-LGs generally estimate a larger mass using the virial theorem. Second, the lower limit of $(\sigma_{r}/\sigma_{t})^{2}$ is 0.51 for the left plot and 0.26 for the right plot. This suggests that removing OLGM's within $350$ kpc from one of the main pairs yields systems with larger tangential velocity and thus more circular orbits around the center of mass of the main pair. When we keep OLGM's with a distance $< 350$ kpc of either the mock-MW or mock-M31, we include satellites that follow the bulk motion of the mock-MW or mock-M31 toward the center of mass resulting in a large radial component. Third, in both plots we see the median (solid-black line) of the VT derived mass tracks the one-to-one (dashed-black) line. This suggests the VT gives a reasonable estimate of the true mass for systems with various true masses.

\section{Impact of the Large Magellanic Cloud} 
\label{sec:LMC}

As mentioned in Section~\ref{sec:Intro}, the TA mass is sensitive to the presence of the LMC. We now investigate how the LMC affects the TA and VT analysis. Given the LMC’s mass  < 20\% MW and close proximity to the MW <50kpc, there are two main effects to consider: 1) a displacement of the Milky way disk from the galaxy’s center of mass, and 2) a shift in the MW-M31 barycenter. The former yields the resulting MW “reflex motion” \citep{2021NatAs...5..251P, 2021ApJ...919..109G}, which alters the true relative velocity of nearby galaxies, including the MW-M31 pair and subsequently the OLGM’s velocity with respect to the barycenter. To calculate the true relative velocity between galaxies in the LG, the reflex motion must be considered by correcting the heliocentric velocities. For 2) the shift in the MW-M31 barycenter affects the calculated OLGM’s line of sight velocity with respect to the barycenter’s location. Since the true relative velocities are easily calculated in the simulation, we can account for the LMC presence by calculating the barycenter shift. Additionally, for all other parameters fixed, the change in the TA mass estimate over the expected range of the LMC mass is $\sim 20\%$~\citep{2016MNRAS.456L..54P}. This systematic is comparable to the level of systematic uncertainty derived in our analysis without including the LMC. Therefore, it is interesting to examine the impact of including LMC-like objects on both our TA and VT analyses.

\subsection{Impact of LMC on Timing Argument Analysis} To  construct our LG sample including LMC-like objects, we have considered two different cuts from our primary sample of LG-like systems.  Our first cut selects LMCs as those galaxies found within 350 kpc from either of the main galaxies in the LG-like system, and requires a dark matter mass cut on LMC-like systems such that ratio of the LMC-like object relative to the main halo mass is $f_c > 0.05$. Given this criteria, we consider this as our most ``relaxed" cut. For our preferred TA cut, corresponding to $V_{t}<|V_{r}|, \,  -195 \, {\rm km \, s}^{-1} < V_{r} < -65 \, {\rm km \, s}^{-1}$, we find 34 LG analogs with an LMC-like object. We refer to this as sample (a) in our discussion in the remainder of this section. Our second cut imposes a stricter requirement to identify LMCs. First, we choose the MW as the fainter of the two galaxies in the main pair and select LMCs as galaxies found within 350 kpc of the mock MW. In addition for this sample, we require $f_{c} > 0.1$. Given these criteria, we consider this as our ``strict" cut. Given our preferred TA sample as above, we find 12 LG analogs with an LMC-like object. We refer to this as sample (b) below. Note that for clarity sample (b) is a subset of sample (a). 

For both of the above samples, we then solve the timing argument twice: once with $V_{r}, V_{t} \, \mathrm{and} \, |\vec{r}|$ calculated when ignoring the LMC presence, and a second time applying the velocity and distance transformations described in~\citet{2016MNRAS.456L..54P} to account for the LMC presence. As a result, we find that the mean and the 90\% containment for the cumulative distributions are similar for both cases. For sample (a) including the effect of the LMS gives a mean of $0.70$ and the 90\% containment regions are $[0.41,1.14]$, while for this same sample not accounting from the LMC gives a mean of $0.73$ and the 90\% containment regions are $[0.38,1.12]$. For sample (b) when accounting for the mean is $0.70$ and the 90\% containment regions are $[0.57,1.04]$, while for this same sample not accounting from the LMC gives a mean of $0.68$ and the 90\% containment regions are $[0.54,1.03]$. These ranges are consistent with those reported in Table~\ref{tab:MTA_Table}. 

\subsection{Impact of LMC on Virial Theorem Analysis}
 \par We now move on to discuss the impact of the LMC on our VT analysis. For this, we consider the same LMC selection cuts (a) and (b) described above. We apply these cuts to our preferred VT sample of OLGMs with $M_{dm} > 10^{9}$ M$_{\odot}$ given in bottom panel of Table~\ref{tab:MVT_Table}. For our cuts (a) defined above, we find 102 analog LGs, while for our cuts (b) defined above we find 26 analog LGs. Again we calculate $M_{VT}$ ignoring the LMC presence and taking into account the presence of the LMC. Here we take the LMC into account by calculation the barycenter of the MW-LMC-M31 and calculate the radial velocity of each OLGM with respect to this new barycenter. We note that we take the MW center of mass coordinate to be the barycenter of the LMC-MW system. As a result, we find that the mean and the 90\% containment for the cumulative distributions are as follows. For sample (a) accounting for the presence of the LMC gives a mean is $0.71$ and the 90\% containment regions are $[0.24,1.43]$, while not accounting for the LMC gives a mean is $0.72$ and the 90\% containment regions are $[0.20,1.02]$. For sample (b) accounting for the effect of the LMC gives a mean of $0.71$ and the 90\% containment regions are $[0.11,1.53]$, while not accounting for the LMC gives a mean of $0.65$ and the 90\% containment regions are $[0.14,1.50]$. 


We note that, given the relatively small numbers of LGs we find in Illustris, it is unlikely to find a ``near exact" LMC-MW-M31 like LG system, indicating how rare the LMC-MW-M31 system is. Nonetheless, it is worthwhile to identify a ``best" M31-MW-LMC candidate system from our cuts above, and examine the TA and the VT for this system. Out of our entire LMC samples (a) and (b) constructed above, we find the ``best" candidate LG system has an LMC-like object 139 kpc from the fainter galaxy in the main pair, and has $f_{c} =0.26$. For this system we do not find a difference in the TA estimate when calculated with or without the presence of the LMC, with $A = 0.68$ 
in both cases. When we calculate the mass using the VT we find a slight change when including the LMC, from $B = 0.85$ to $B_{LMC} = 0.73$. In both cases, these values are within 10\% of the median value in Table~\ref{tab:MVT_Table}, which is calculated for the entire sample of LG-like systems.

\section{Discussion and Conclusions}
\label{sec:discussion}

\par We have identified a sample of Local Group-like systems in the IllustrisTNG simulation, and have used this sample to study two mass estimators for the Local Group: the timing argument and the virial theorem. For the timing argument mass, we update both the slight bias in this estimator and the cosmic scatter, including measurements of the cosmological constant and the Milky Way-M31 tangential velocity. Though our primary Local Group-sample is defined using an absolute magnitude cut in addition to kinematic cuts, compared to previous calculations which identified the Local Group-sample based on dark matter halos masses, our results are in good agreement~\citep{2008MNRAS.384.1459L,2017MNRAS.468.1300P}. We find that the timing argument slightly overestimates the Local Group mass, though this estimator is unbiased at the $90\%$ c.l. This result depends weakly on the precise kinematic cuts that we use to define the Local Group-like sample. 

\par Our results are also consistent with previous results in that we find that the cosmic scatter in determining the timing argument mass is at least as significant as the statistical uncertainty. In particular, for the timing argument we find that the Local Group mass at 90\% c.l. is $4.75_{-2.41}^{+2.22} \times 10^{12}$ M$_\odot$, where the errors are due to the cosmic scatter. For comparison, the statistical uncertainty quoted in~\citet{2012ApJ...753....8V} is $0.45 \times 10^{12}$ M$_\odot$. At this time, the largest uncertainty in the timing argument mass comes from the tangential velocity of M31. In our analysis, we use the updated the measurement from Gaia EDR3, which gives a tangential velocity of $79$ km/s, larger than the previously determined value of $17$ km/s from HST. For the radial velocity and the MW-M31 separation fixed, going from a tangential velocity of $17$ km/s to $79$ km/s increased the timing argument mass from $5 \times 10^{12}$ M$_\odot$ to $6 \times 10^{12}$ M$_\odot$. 


\par We have provided the first comparison of the Local Group mass as determined from the virial theorem to that obtained in simulations of Local Group-like systems. We find that the virial theorem estimator also overestimates the true Local Group mass, though there is a larger scatter in the virial mass to true mass ratio relative to the corresponding ratio for the timing argument. For the virial theorem, we find the Local Group mass at 90\% c.l. is $2.00_{-1.5}^{+2.1} \times 10^{12}$ M$_\odot$, where the errors are due to the cosmic scatter. For comparison, the statistical uncertainty quoted in~\citet{2014MNRAS.443.1688D} is $0.4 \times 10^{12}$ M$_\odot$. So similar to the timing argument, the cosmic scatter in this mass estimator dominates the uncertainty budget.

\par We attribute the slight bias and broad scatter in the virial theorem estimator to several factors. First, we find that the orbits are predominantly radial for Local Group galaxies, which differs from the virial theorem assumption of isotropic orbits. In particular, we find that the ratio of radial to tangential orbits for Local Group dwarf galaxies is $\sigma_r^2/\sigma_t^2 \simeq 2$, implying that these galaxies are on largely radial orbits. This small amount of tangential motion is likely to not only impact the virial theorem mass estimator, but also other estimates of the Local Group mass based on pure radial infalling motion~\citep{2016MNRAS.456L..54P}.

 It is worth noting the $A$ and $B$ distributions presented in this paper depend on the mass definition used as the true galaxy mass in the simulation. In particular, there are two mass definitions to consider: $M_{200}$ is the total mass found within the virial radius of the galaxy, and $M_{halo}$ is the total mass of all particles bound to the halo. Since the $M_{halo}$ definition includes mass outside of the virial radius, this is typically larger than $M_{200}$ for a given halo.~\cite{2008MNRAS.384.1459L} investigated how the timing argument depends on which mass definition is used by calculating the distribution of $A$ twice, once for each mass definition. First, the for their main analysis these authors calculate $A$ assuming $M_{200}$ mass, then a second time where they use $M_{halo}$ as the true mass to calculate $A_{halo}$. These authors find that while the distribution of $A$ remains unchanged, the entire distribution of $A_{halo}$ is shifted up by 16-20\% from $A$. This result is expected, since $M_{halo}$ is typically larger than $M_{200}$. The mass definition used in our paper corresponds to $M_{halo}$ in ~\cite{2008MNRAS.384.1459L}, thus our distribution of $A$ should be compared to their $A_{halo}$ distribution. Assuming that we instead used the $M_{200}$ mass definition, our resulting distribution of $A$ and $B$ would be shifted down by 16-20\%.  

 In this paper we attribute the largest source of uncertainty in the virial theorem and timing argument mass estimates to cosmic scatter. There are several physical effects that could contribute to the cosmic scatter. Two possible physical effects are: 1) the larger scale density effects in the cosmic web, and/or 2) galaxy merger history. We can understand the former through the varying local gravitational potential, which depends on each mock-LG's location in cosmic web - whether it resides in clusters, filaments, or voids that can significantly affect the kinematics of nearby galaxies (satellites and OLGM’s). In particular, \citet{2020ApJ...900..129W} find the kinematics of satellite galaxies and OLGM’s show a strong correlation to the hosts location in the large-scale structure (LSS). Since the virial theorem uses the kinematics of OLGMs, we predict the mock-LGs location in the LSS introduces significantly more cosmic scatter in the virial theorem than the timing argument.  
Another possible physical affect that can contribute to the cosmic scatter is Galaxy merger history.  One could attempt to uncover whether differing galaxy merger histories introduce cosmic scatter by splitting our sample based on merger history and plotting the distribution of $A$ and $B$ for these samples.
\par For both the timing argument and the virial theorem, we have provided an estimate of the impact of LMC-like systems on the analysis. Though we are not able to identify an exact LG-like system with a MW-M31-LMC combination, for candidate systems with LMCs we find the timing argument and the virial theorem are slightly shifted such that the LG mass is overestimated in these analyses. However, this shift occurs both with and without accounting for the dynamical presence of the LMC. This preliminary analysis indicates that for systems with LMC-like galaxies, the LG masses are shifted by $\lesssim 10\%$, though there is still substantial scatter in the results. We do not find exact LMC-MW-M31 analogues in the Illustris simulations. However, in LG simulations with satellite companions as massive as the LMC we find that the effect on the TA and VT estimators is small, though a more in-depth study is likely required on a larger sample of LMC-MW-M31 systems to confirm these results. Indeed, since the LMC-MW-M31 system appears to be a rare outlier, it may be that a more detailed understanding of the statistical distribution of LG-like systems is required in order to robustly derive the LG mass as we have done in Section~\ref{sec:results}.


\par More generally, the radial and tangential motions of the dwarf satellites have important implications for understanding the dynamics of dwarf galaxies in the Local Group. Though the current measurements of the proper motions of isolated dwarfs are not precise enough~\citep{2021MNRAS.501.2363M,2021arXiv210608819B} to be constrained by our predictions, future Gaia data releases may be able to improve upon the astrometry of the isolated dwarfs. 

\par An additional source of uncertainty in our virial theorem estimator may lie in our choice of the outermost radius that defines the extent of the Local Group. For all Local Group analogs in Illustris, we fix this radius at 1.5 Mpc. The corresponds to the approximate turn-around radius of the Local Group, which is about a factor of two greater than the simple estimate for the Local Group virial radius based upon the dynamical free-fall time. We have checked that the true mass to virial mass distribution is unaffected by the precise choice of outer radius for the Local Group. Indeed, defining 0.7 Mpc as the outer radius does not change our statistical distribution of $B$ derived in Section~\ref{sec:results}, though it does cut down on the sample of galaxies that can be used in the virial analysis. In addition, the ratio $\sigma_r^2/\sigma_t^2 \simeq 2$ remains similar for this sample. This indicates that the deduced mass and the kinematics of the Local Group is independent of where the exact outer radius is set, and that the cosmic scatter may reflect more simplifying assumptions in the underlying nature of the equilibrium of the Local Group.

\section*{Acknowledgements}
LES acknowledges support from DOE Grant de-sc0010813. We are grateful to Alexander Riley for many engaging conversations on this paper. This work was supported by a Development Fellowship from the Texas A$\&$M University System National Laboratories Office. This work was supported by NSF grant AST-1263034, “REU Site: Astronomical Research and Instrumentation at Texas A\&M
University.” 

\section*{Data Availability}
The data underlying this article will be shared on reasonable request to the corresponding author.

\bibliographystyle{mnras}

\bsp	
\label{lastpage}
\end{document}